\documentclass[pra,aps,showpacs,twocolumn]{revtex4}

\usepackage{amsmath,amsfonts,amssymb,caption,hyperref,color,epsfig,graphics,subfigure,graphicx,latexsym,mathrsfs,placeins,theorem,url,verbatim,epstopdf,float,adjustbox}
\usepackage{caption}
\usepackage{booktabs}
\usepackage{array}
\usepackage{afterpage}
\usepackage{bbm}

\usepackage{algorithm}
\usepackage{algorithmic}

\makeatletter
\renewcommand{\fs@ruled}{\def\@fs@cfont{\bfseries}\let\@fs@capt\floatc@ruled
  \def\@fs@pre{\hrule height1.2pt depth0pt \kern2pt}%
  \def\@fs@post{\kern2pt\hrule height0.4pt depth0pt \let\@fs@capt\floatc@ruled}%
  \def\@fs@mid{\kern2pt\hrule height0.4pt depth0pt \kern2pt}%
  \let\@fs@iftopcapt\iftrue}
\makeatother
\floatstyle{ruled}
\restylefloat{algorithm}

\allowdisplaybreaks[3]

\hypersetup{colorlinks,linkcolor={blue},citecolor={blue},urlcolor={black},pdfborder={0 0 0}}

\makeatletter

\newcommand{\Rmnum}[1]{\expandafter\@slowromancap\romannumeral #1@}
\makeatother

\newtheorem{definition}{Definition}
\newtheorem{lemma}[definition]{Lemma}

\newtheorem{theorem}{Theorem}
\newtheorem{corollary}[definition]{Corollary}

\newcommand{\Tr}{\mathrm{Tr}}
\newcommand{\abs}[1]{\left|#1\right|}
\newcommand{\norm}[1]{\left\|#1\right\|}
\newcommand{\bra}[1]{\langle#1|}
\newcommand{\ket}[1]{|#1\rangle}

\newcommand{\E}{\mathbb{E}}
\newcommand{\Var}{\mathbb{V}}
\newcommand{\rev}[1]{#1}

\def\squareforqed{\hbox{\rlap{$\sqcap$}$\sqcup$}}
\def\qed{\ifmmode\squareforqed\else{\unskip\nobreak\hfil
        \penalty50\hskip1em\null\nobreak\hfil\squareforqed
        \parfillskip=0pt\finalhyphendemerits=0\endgraf}\fi}
\newenvironment{proof}{\noindent \textbf{{Proof.~} }}{\qed}
\def\bpf{\begin{proof}}
\def\epf{\end{proof}}

\begin{document}

\title{Estimating the Quantum Hockey--Stick Divergence on Near-Term Quantum Hardware}

\author{Mingyu Liu}\email[]
{2024210568@buct.edu.cn}
\affiliation{College of Information Science and Technology,
    Beijing University of Chemical Technology, Beijing 100029, China}
\author{Xian Shi}\email[]
{shixian01@gmail.com (corresponding author)}
\affiliation{College of Information Science and Technology,
    Beijing University of Chemical Technology, Beijing 100029, China}

\date{\today}

\begin{abstract}
The quantum Hockey--stick divergence quantifies how well two
quantum states can be distinguished. It plays a central role in many tasks of quantum
information. Estimating this nonlinear spectral functional from copies of unknown
states remains difficult on near-term hardware. Existing methods either need exponentially many samples, require deep fault-tolerant
circuits or oracles unavailable from copies, or rely on heuristic training without convergence or accuracy guarantees.
To address these problems, we propose the regularized
Monte-Carlo method on shallow circuits (RMSC), a divergence estimator
that matches current frontier devices in circuit depth and gate fidelity
at practically relevant auditing precisions. Its sample complexity scales quasi-quadratically with the inverse precision and
is independent of the number of qubits. Next, we provide a complete error analysis and remove the regularization bias via purely classical post-processing, restoring the
true divergence. Finally, a
complete quantum differential privacy audit validates the estimator.
\end{abstract}

\maketitle

\section{Introduction}\label{sec:intro}

In classical information theory, the Hockey--stick divergence is a
fundamental member of the $f$-divergence family generated by
the function $f_t(x)=\max\{0,x-t\}$ ($t\ge0$) \cite{polyanskiy10}.
It directly captures the optimal error trade-off in binary
hypothesis testing under a threshold weight $t$ \cite{polyanskiy10}.
Extending
this concept to the quantum domain, Sharma and Warsi introduced the
quantum Hockey--stick divergence as a generalization of the trace
distance to establish tight converse bounds for one-shot quantum
hypothesis testing and quantum channel coding \cite{sharma13}. Beyond state discrimination and communication, this divergence is
widely used in quantum resource theories~\cite{lami25}, entropy
continuity bounds~\cite{winter16}, quantum differential privacy
(QDP)~\cite{hrf23,zy17,ar19,kairouz17}, and integral representations of
quantum entropies and divergences~\cite{frenkel23,hirche24}. Estimating the quantum Hockey--stick
divergence from copies of quantum states is therefore a basic
primitive for all these tasks.

Present-day noisy intermediate-scale quantum (NISQ) devices run
shallow circuits with cheap, repeatable measurements, but lack
error correction, so the usable depth and gate fidelity are
limited~\cite{preskill18}. On these devices, ground-state energies,
local observables, and state fidelities are estimated
routinely~\cite{peruzzo14,kandala17}. In contrast, estimating the
Hockey--stick divergence for general states remains difficult.
Because it sums the positive eigenvalues of $H=\rho-t\sigma$, linear
observables cannot extract it directly. Existing estimators
therefore face clear limits on near-term hardware. State tomography and classical shadows need only copies, and shadows
predict many linear observables with shallow measurements. Nevertheless,
for global nonlinear spectral functionals, both require samples
exponential in the qubit number $n$~\cite{hhw16,hkp20,cchl21}. Quantum singular value transformation (QSVT) achieves
polynomial query complexity for such spectral functionals, but
requires a block-encoding of $H$, an oracle that state copies alone
cannot provide~\cite{gil19}. Later frameworks recompile these query
algorithms to run on state copies. Yet the resulting circuits
remain deep and fault-tolerant, with costs polynomial in the rank
and precision, or in the Hilbert-space dimension
$d=2^n$~\cite{wang24fast,wang25samplizer}. Phase-estimation schemes work directly from copies, but need long
coherent circuits, with costs growing as a high power of $n$
or exponentially in the precision~\cite{ghosh26,berg20}. Variational and quantum neural estimators use parameterized
shallow circuits suited to near-term
devices~\cite{chen22vtde,goldfeld24,lu26}. However, their heuristic
training can suffer from barren
plateaus and mostly lacks convergence or accuracy
guarantees~\cite{mcclean18}. Even recent quantum neural estimators with theoretical
bounds need, in general, samples scaling linearly with
$d=2^n$~\cite{sreekumar26}. Taken together, these estimators either
need exponentially many samples or deep fault-tolerant circuits, or they
lack analytical guarantees. A paradigm better suited to these restrictions is Monte-Carlo sampling~\cite{metropolis49}, which decomposes
one complex estimation task into many independent, simple tasks
by the law of large numbers.
This idea is well tested in sampling-based Hamiltonian simulation~\cite{campbell19,berry15},
in measurement-based computation~\cite{raussendorf01}, and in quantum Metropolis
sampling~\cite{temme11}. In
the extreme, the logical depth is $O(1)$ and the sample count
becomes exponential \cite{hhw16}. For the Hockey--stick
divergence we address the following question: \emph{At what
sample complexity can the divergence be estimated from copies by
circuits of polynomial depth, with certified accuracy?}

In this manuscript, we propose the regularized Monte-Carlo method
on shallow circuits (RMSC) to estimate the Hockey--stick
divergence from state copies. Rather than evaluating the non-smooth
functional $E_t$ directly, RMSC estimates a spectrally regularized
surrogate $E^{(\eta)}_t$ using shallow Hadamard-test circuits
executed on fresh copies, achieving a qubit-independent sample
complexity of $\widetilde O(\varepsilon^{-2}\ln^{2}(1/\varepsilon))$
at target accuracy $\varepsilon$ (Theorem~\ref{thm:perf}). Although
the spectral smoothing introduces a deterministic certification gap
$G_\eta = E_t - E^{(\eta)}_t$, Theorem~\ref{thm:main} characterizes
$G_\eta$ in closed form, and Theorem~\ref{thm:inv} eliminates this
regularization bias via purely classical post-processing---specifically,
by analytically inverting the forward bias map---to yield a compensated
point estimate whose expectation never exceeds $E_t$. Section~\ref{sec:experiments}
validates this complete estimation and compensation pipeline on three
QDP channels using only copies of their output states. Crucially,
repeating shallow circuits is well suited to near-term processors: after
optimization, at precision $\varepsilon\approx0.1$ with
$N\approx3\times10^{3}$ samples per side, RMSC requires
a per-gate error rate of about $1.7\times10^{-3}$ and a typical
circuit depth of about $10^{2}$ gate layers. This depth corresponds to
roughly $10$~ms per sample on trapped-ion hardware (two-qubit
gates of order $0.1$~ms) and a few microseconds on superconducting
hardware (gates of $30$--$90$~ns). By comparison, frontier
trapped-ion processors achieve two-qubit error rates down to
$1.84\times10^{-3}$ with coherence times
exceeding the per-sample duration by orders of magnitude~\cite{moses23}, while
frontier superconducting processors report error rates around
$3.3\times10^{-3}$ with coherence times
$T_{1}\approx68$--$98~\mu$s that cover each
shallow execution several times over~\cite{acharya25}. The resource requirements
of RMSC thus match the capabilities of frontier NISQ devices. To
our knowledge, \emph{this is the first estimator of the Hockey--stick
divergence with this property}.

The paper is organized as follows. Section~\ref{sec:prelim} fixes the notation;
Section~\ref{sec:algorithm} presents RMSC and its three guarantees (sample complexity,
regularization gap, and inversion compensation); Section~\ref{sec:perf} compares
RMSC with prior approaches and assesses its NISQ resource envelope;
Section~\ref{sec:experiments} reports the numerical experiments; and
Section~\ref{sec:conclusion} concludes.

\section{Preliminaries}\label{sec:prelim}

We first fix the notation, then recall the definition and the
properties of the quantum Hockey--stick divergence used later.

\emph{\rev{Notation.}} 
Two parameters control the estimation. The precision
$\varepsilon\in(0,1/2)$ is the target accuracy of the estimator.
The threshold $t=e^{\epsilon_{\mathrm{DP}}}\ge1$ is set by the privacy
parameter $\epsilon_{\mathrm{DP}}$ of QDP; the two symbols are
unrelated. The symbol $\norm{\cdot}$ is the operator norm.
The eigenvalues of a density matrix lie in $[0,1]$, so the
triangle inequality gives
$\norm{H}\le\norm{\rho}+t\norm{\sigma}\le1+t$; this bound is used
throughout. The expectation $\E[\cdot]$ averages over two sources
of randomness, the sampled variable $y$ and the measurement
outcomes. We write
$f=\Theta(g)$ if $c_1 g\le f\le c_2 g$ for constants $c_1,c_2>0$,
and $\widetilde O$ hides factors polylogarithmic in $1/\varepsilon$.

\emph{\rev{Quantum state divergences.}} The quantum Hockey--stick divergence belongs to the classical
$f$-divergence family \cite{ali66}: it is obtained from the convex function $f(x)=(x-\gamma)_+$,
and its quantum extension is the quasi-entropy of Petz \cite{petz86}.
For states $\rho$ and $\sigma$ and a threshold $\gamma\ge1$ it is defined as
\begin{equation}\label{eq:defhs}
E_\gamma(\rho\|\sigma):=\Tr(\rho-\gamma\sigma)_+,
\end{equation} where $X_+$ denotes the positive part of a Hermitian
operator $X$ in its spectral decomposition \cite{wat18}.
The divergence is nonnegative; it vanishes if and only if $\rho=\sigma$ and $\gamma=1$;
it is jointly convex in $(\rho,\sigma)$; and it obeys the data-processing inequality
$E_\gamma(\Lambda(\rho)\|\Lambda(\sigma))\le E_\gamma(\rho\|\sigma)$
for every quantum channel $\Lambda$ (a completely positive trace-preserving map) \cite{wat18}.
These properties follow from the variational characterization
\begin{equation}\label{eq:varform}
E_\gamma(\rho\|\sigma)=\max_{0\le M\le I}\Tr\bigl(M(\rho-\gamma\sigma)\bigr),
\end{equation} where the maximum is attained at the spectral projector $M=\Pi_+=\mathbbm{1}_{\lambda(\rho-\gamma\sigma)>0}$.
The proofs are collected in Appendix~\ref{app:ht}.

\emph{\rev{Spectral form.}} \rev{Let $H=\rho-t\sigma$ and let
$\Pi_+=\mathbbm{1}_{\lambda(H)>0}$ denote the spectral projector onto the
positive eigenspace of $H$, with $\lambda(H)$ the spectrum of $H$. By
spectral calculus,
\begin{align}\label{eq:spectral}
E_t(\rho\|\sigma)&=\Tr\bigl((\rho-t\sigma)\Pi_+\bigr)
=\omega_\rho-t\,\omega_\sigma,\\
\omega_\rho&:=\Tr(\rho\Pi_+),\qquad \omega_\sigma:=\Tr(\sigma\Pi_+).
\end{align}
Estimating the divergence therefore reduces to estimating the two spectral
weights $\omega_\rho$ and $\omega_\sigma$. The latter is obtained by the same
circuits with the working register initialized in $\sigma$; the Hamiltonian
$H=\rho-t\sigma$ is kept unchanged.}

\emph{\rev{Spectrally regularized divergence.}}
RMSC does not estimate $E_t$ directly. It estimates a
regularized version of it, defined as follows. Let
$\operatorname{erf}(x)=\frac{2}{\sqrt\pi}\int_0^x e^{-u^2}\,du$ be the error
function, and let $\operatorname{erfc}=1-\operatorname{erf}$ be its
complement. Both act on Hermitian matrices by functional
calculus. For $\eta>0$ define the \emph{spectrally regularized
divergence}
\begin{equation}\label{eq:regdiv}
E^{(\eta)}_t(\rho\|\sigma):=\sum_k g_\eta(\lambda_k),
\qquad
g_\eta(\lambda):=\frac{\lambda}{2}\Bigl(1+\operatorname{erf}
\tfrac{\lambda}{2\sqrt\eta}\Bigr),
\end{equation}
where $H=\rho-t\sigma=\sum_k\lambda_k\ket{k}\bra{k}$ as above.
The function $g_\eta$ replaces the positive part
$\lambda_+=\max(\lambda,0)$ by a smooth ramp of width
$2\sqrt\eta$. Eigenvalues far from zero, $|\lambda|\gg2\sqrt\eta$,
contribute to $E^{(\eta)}_t$ exactly as they contribute to $E_t$.
Eigenvalues close to zero, $|\lambda|\lesssim2\sqrt\eta$, contribute
about half of their magnitude. The difference between $E_t$ and
$E^{(\eta)}_t$ is called the certification gap $G_\eta$. Theorem~\ref{thm:main}
of Section~\ref{sec:algorithm} gives this gap exactly.

\section{Algorithm and main theorems}\label{sec:algorithm}

\rev{This section presents RMSC together with its three
guarantees.} Theorem~\ref{thm:perf} gives the sample complexity
of the estimator and the hardware noise it tolerates. \rev{Theorem~\ref{thm:main} relates
the target $E^{(\eta)}_t$ and the true divergence $E_t$ through the
certification gap $G_\eta$. Theorem~\ref{thm:inv} and Corollaries~\ref{cor:cert}
and~\ref{cor:rec} supply the compensated point estimate and the safe
lower bound used in the QDP application.}
\begin{theorem}[performance guarantee]\label{thm:perf}
\rev{Fix an estimation precision $0<\varepsilon<1/2$, a failure
probability $\beta\in(0,1)$, and the regularization parameter
$\eta>0$ of the target $E^{(\eta)}_t$. For $N$ Monte-Carlo samples
per side, there exists a hybrid quantum--classical
estimator that accesses the two states only through independent
copies of $\rho$ and $\sigma$ and outputs an estimator
$\hat E_t=\hat\omega_\rho-t\,\hat\omega_\sigma$ satisfying
\begin{equation}\label{eq:thmN}
\Pr\bigl[\abs{\hat E_t-E^{(\eta)}_t}\le\varepsilon\bigr]\;\ge\;1-\beta,
\qquad
N=\widetilde O\!\left(\varepsilon^{-2}\ln^2\frac{1}{\varepsilon}\right),
\end{equation}
provided that each of its samples, a circuit of typical depth
$a=O\bigl(\eta^{-3/4}\varepsilon^{-1/2}\bigr)$ gate layers (for fixed $t$), runs on
hardware whose per-gate error rate does not exceed
$b=\widetilde O(\varepsilon^{3})$.}
\end{theorem}

\rev{RMSC is specified below (Algorithm~\ref{alg:main}); the
theorem is proved in Appendix~\ref{app:thm2}.} At precision
$\varepsilon\approx0.1$ they are $a\approx10^{2}$ gate layers and
$b\approx1.7\times10^{-3}$. The noise budget behind $b$ is
derived in Appendix~\ref{app:resnoise}, and
Section~\ref{sec:nisq} compares both values with current frontier
devices. The sample count $N$ depends on the target precision
$\varepsilon$ alone. The count is set by the Monte-Carlo statistical error, which
by the law of large numbers scales as $1/\sqrt{N}$
regardless of the Hilbert-space dimension. However, this carries an implicit
message: as the number of qubits of the states to be measured
grows, the per-sample circuit depth does not increase, but the
parallel circuit width increases. The resulting difficulty of
circuit implementation, such as crosstalk noise \cite{sarovar20},
will be an open problem that needs to be further addressed.

\rev{We now describe the RMSC estimator that realizes this
guarantee. It estimates the spectral weight $\omega_\rho$ in three steps.}

\emph{\rev{Step 1: integral representation and regularization.}} \rev{Up to
the nullspace of $H$ (which contributes zero to $\Tr(H\Pi_+)$), the
spectral projector $\Pi_+$ acts as the matrix step function of $H$ and
admits the Fourier integral representation
\begin{equation}\label{eq:fourier}
\Pi_+=\frac{I}{2}+\frac{1}{\pi}\int_0^\infty \frac{\sin(Hy)}{y}\,dy.
\end{equation}
Substituting Eq.~(\ref{eq:fourier}) into $\omega_\rho=\Tr(\rho\Pi_+)$ yields an
improper oscillatory integral; inserting the Gaussian convergence factor
$e^{-\eta y^2}$ ($\eta>0$) regularizes the large-$y$ tail and gives
\begin{equation}\label{eq:Eeta}
\omega_{\rho,\eta}=\frac{1}{2}+\frac{1}{\pi}\int_0^\infty
\frac{e^{-\eta y^2}}{y}\,\Tr\bigl(\rho\sin(Hy)\bigr)\,dy.
\end{equation}
The complete derivation of Eqs.~(\ref{eq:fourier}) and (\ref{eq:Eeta}),
together with the proof that the combined regularized weights
$\omega_{\rho,\eta}-t\,\omega_{\sigma,\eta}$ equal the spectrally
regularized divergence $E^{(\eta)}_t(\rho\|\sigma)$ of Eq.~(\ref{eq:regdiv}),
is provided in Appendix~\ref{app:aux} (Lemmas~\ref{lem:fourier}
and~\ref{lem:reg}). Combining the resulting regularization bias across both
states yields the exact certification gap $G_\eta$ characterized in
Theorem~\ref{thm:main}.}

\emph{\rev{Step 2: Monte Carlo reformulation with importance sampling.}}
\rev{The integral (\ref{eq:Eeta}) can be evaluated by Monte-Carlo integration
with importance sampling, the standard variance-reduction
technique that draws $y$ from a chosen density and reweights the
samples \cite{kahn53}, over any sampling density supported on $[\alpha,\infty)$. The naive choice,
the half-normal $P(y)=2\sqrt{\eta/\pi}\,e^{-\eta y^2}$, makes the estimator
weight $1/y$ diverge at small $y$: the single-shot readout $\hat S\in\{\pm1\}$
with $\E[\hat S]=S(y)=\Tr(\rho\sin(Hy))$ yields a per-sample variance
$\propto1/y^2$, and the total variance carries a $\Theta(1/\alpha)$ penalty
that forces an unfavorable trade-off between the cutoff $\alpha$ and the
sample count $N$. We instead sample from the importance density
\begin{align}\label{eq:Q}
Q(y)&=\frac{e^{-\eta y^2}}{y\,Z}\quad\text{on }[\alpha,\infty),\nonumber\\
Z&=\int_\alpha^\infty\frac{e^{-\eta y^2}}{y}\,dy=\tfrac12 E_1(\eta\alpha^2),
\end{align}
with $E_1$ the exponential integral. The key observation is that the weight
$1/y$ is then absorbed by the density: with the single-sample random variable
\begin{equation}\label{eq:xhat}
\hat{X}(y)=\frac{Z}{\pi}\,\hat S(y),
\end{equation}
we have
\begin{align}\label{eq:unbiased}
\E_Q[\hat X]&=\frac{Z}{\pi}\int_\alpha^\infty Q(y)\,S(y)\,dy\nonumber\\
&=\frac{1}{\pi}\int_\alpha^\infty\frac{e^{-\eta y^2}}{y}S(y)\,dy
=\omega_{\rho,\eta}-\frac{1}{2},
\end{align}
so $\hat X$ is an unbiased estimator of the regularized spectral weight, and
its variance satisfies
\begin{equation}\label{eq:isvar}
\Var[\hat X]\le (Z/\pi)^2
=O\bigl(\ln^2(1/\eta\alpha^{2})\bigr),
\end{equation}
independent of $y$: the $1/y^2$ blow-up is eliminated by construction. The
samples are drawn classically by the inverse cumulative-distribution-function
(inverse-CDF) method \cite{devroye86}; the implementation is given in
Appendix~\ref{app:isimpl}.}

\emph{\rev{Step 3: quantum circuits.}} \rev{The region
$y<\alpha$ is not sampled at all: its contribution to $\omega_{\rho,\eta}$ is
evaluated analytically. Using the first-order Taylor expansion
$\Tr(\rho\sin(Hy))/y\approx\Tr(\rho H)$,
\begin{equation}\label{eq:smallcorr}
\frac{1}{\pi}\int_0^\alpha\frac{e^{-\eta y^2}}{y}\Tr(\rho\sin(Hy))\,dy
\;\simeq\;c_s\,\Tr(\rho H),
\end{equation}
with $c_s=\mathrm{erf}(\alpha\sqrt\eta)/2\sqrt{\pi\eta}$,
up to a bias $O(\alpha^3)$ (Lemma~\ref{lem:trunc}). The quantity
$\Tr(\rho H)=\Tr(\rho^2)-t\Tr(\rho\sigma)$ is estimated once by constant-depth
SWAP tests \cite{nc10}; no sampling is performed in $[0,\alpha)$, and the
estimator (\ref{eq:xhat}) is therefore bounded for every $y\ge\alpha$. The
small-time shot-noise divergence is removed analytically rather than
truncated statistically.}

\rev{For $y_i\ge\alpha$, the signal
$S(y_i)=\Tr(\rho\sin(Hy_i))$, the imaginary part of
$\Tr(\rho e^{iHy_i})$, is extracted by a Hadamard test whose
controlled unitary $e^{iHy_i}$ is implemented with the LMR
density-matrix exponentiation protocol \cite{lmr14}: the evolution is
compiled into Trotter steps, i.e.\ factors of the first-order product
formula $e^{iHy}\approx(e^{i\rho\Delta y}\,e^{-it\sigma\Delta y})
^{y/\Delta y}$, and each step consumes one fresh copy of $\rho$
(respectively $\sigma$); LMR requires neither preparation circuits nor
block-encoding oracles, matching the copies-only setting exactly. We
additionally find a zero-cost accuracy improvement here: at the trace
level the first-order Trotter bias is one order smaller than its
operator-norm bound, because the leading error term is proportional to
the commutator $[\rho,\sigma]$ and
$\Tr(\rho[\rho,\sigma])=\Tr(\sigma[\rho,\sigma])=0$
(Appendix~\ref{app:trace}). The Hadamard-test and LMR circuit
identities are collected in Appendix~\ref{app:circ}, and the Trotter
error bounds in Appendix~\ref{app:aux}.}

\emph{\rev{Classical post-processing.}} \rev{For each sample, the classical
computer forms $\hat{X}_i=(Z/\pi)\hat S_i$ and averages; adding the analytic
correction (\ref{eq:smallcorr}),
\begin{equation}\label{eq:omega_rho}
\hat\omega_\rho=\frac{1}{2}+c_s\,\Tr(\rho H)+\frac{1}{N}\sum_{i=1}^N\hat{X}_i.
\end{equation}
Running the same procedure with the working register initialized in $\sigma$
gives $\hat\omega_\sigma$ ($H$ unchanged). Equivalently, the divergence can
be obtained from a single additional run with the parameter inverted, via the
duality relation
\begin{equation}\label{eq:duality}
E_t(\rho\|\sigma)=1-t+t\,E_{1/t}(\sigma\|\rho).
\end{equation}
The final estimate is assembled as
\begin{equation}\label{eq:final}
\widehat E_t=\hat\omega_\rho-t\,\hat\omega_\sigma.
\end{equation}
The complete procedure is summarized in Algorithm~\ref{alg:main}.}

\begin{figure*}[t]
\centering
\includegraphics[width=\textwidth]{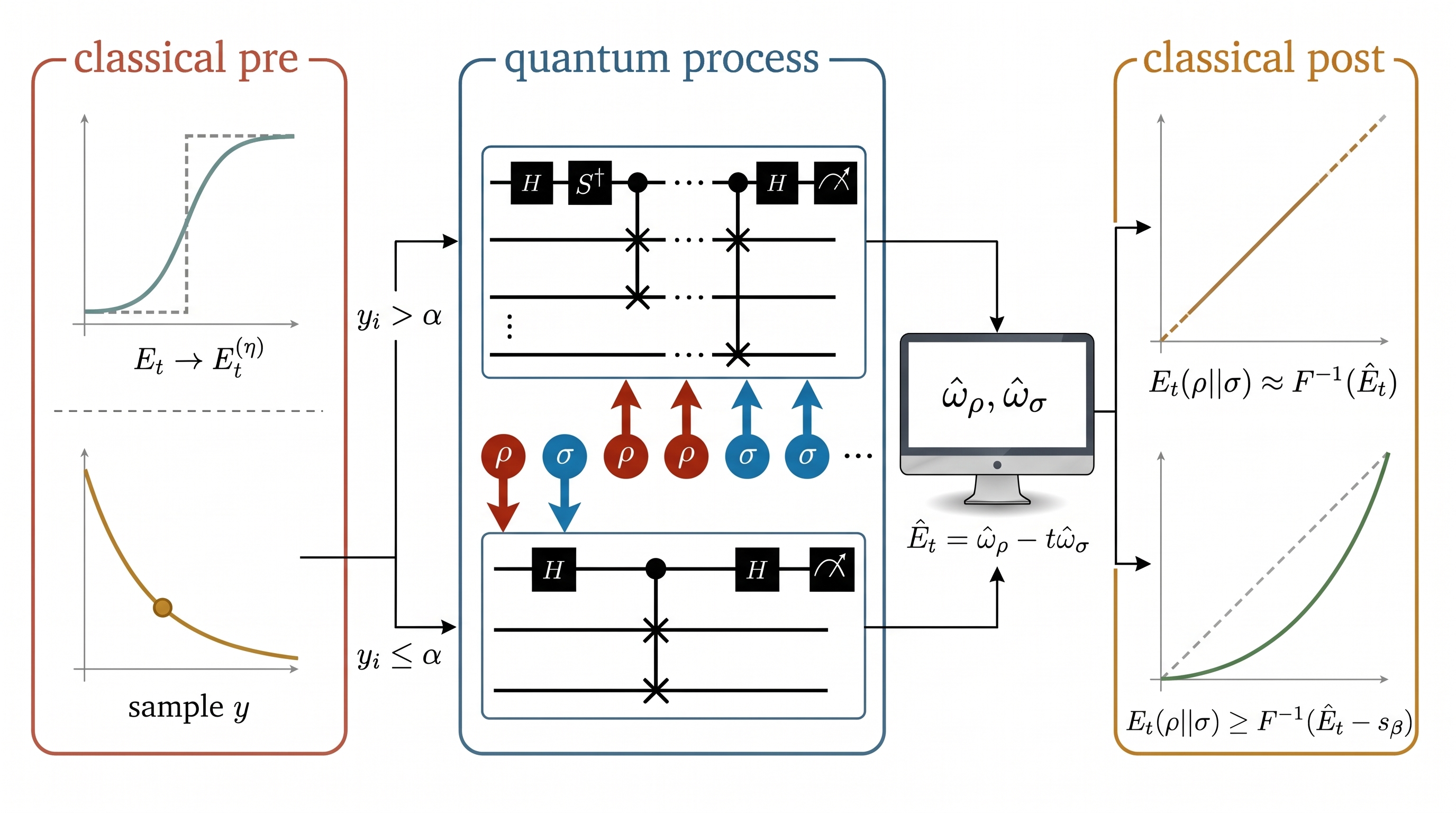}
\caption{Estimating the Hockey--stick divergence with RMSC. The original problem is converted into a step function. Regularization is applied and sampling is performed, and each sampling result is taken as the evolution time of a quantum circuit. The short-time region is not sampled but treated analytically, directly with a SWAP test. Finally, the error of the regularization is compensated through the inverse function. Here $\times$ denotes the exponentiated SWAP gate $e^{i\tilde{S}\Delta y}$ in the upper circuit, where $\tilde{S}$ denotes the SWAP gate and $\Delta y$ is the angle; the lower circuit is a controlled SWAP. $S^\dagger$ is the adjoint of the phase gate $S=\mathrm{diag}(1,i)$.}
\label{fig:flow}
\end{figure*}

\emph{\rev{Shallow per-sample circuits.}} \rev{To summarize the flow: each
sample is read out by a shallow circuit: the quantum processor
evaluates one per-sample observable on a fresh copy, and the classical
machine
performs all sampling, averaging, and error budgeting. $N$ independent
shallow reads thus replace one deep evolution: compared with
phase-estimation schemes, whose depth grows exponentially with $n$, and
with tomography, whose per-copy circuits are $O(1)$ but which require
exponentially many copies, RMSC occupies the middle ground of a
moderate, $n$-free per-sample depth paid in repetition. Figure~\ref{fig:flow} summarizes the workflow.}

\begin{algorithm}[htbp]
\caption{RMSC estimation of the Hockey--stick divergence}
\label{alg:main}
\begin{algorithmic}[1]
\REQUIRE \textbf{Input:} copies of $\rho,\sigma$; $t\ge1$; $\eta,\alpha>0$; $N$
\ENSURE \textbf{Output:} estimate $\widehat E_t$ of $E_t(\rho\|\sigma)$
\STATE compute the correction coefficient $c_s=\mathrm{erf}(\alpha\sqrt\eta)/(2\sqrt{\pi\eta})$ and the normalization $Z=\tfrac12E_1(\eta\alpha^2)$
\STATE sample $y_1,\dots,y_N$ from the importance density $Q(y)=e^{-\eta y^2}/(yZ)$ on $[\alpha,\infty)$
\FOR{$i=1$ to $N$}
\STATE run the Hadamard test with LMR evolution to obtain $\hat{S}_i\in\{\pm1\}$, $\E[\hat S_i]=S(y_i)$
\STATE set $\hat{X}_i=(Z/\pi)\,\hat{S}_i$
\ENDFOR
\STATE estimate $\Tr(\rho H)$ and $\Tr(\sigma H)$ by SWAP tests
\STATE $\hat\omega_\rho=\frac12+c_s\,\Tr(\rho H)+\frac{1}{N}\sum_i\hat{X}_i$
\STATE repeat lines 2--7 with $\sigma$ as the input state ($H$ unchanged; the constants $c_s$ and $Z$ of line 1 are state-independent) to obtain $\hat\omega_\sigma$
\STATE $\widehat E_t=\hat\omega_\rho-t\,\hat\omega_\sigma$
\end{algorithmic}
\end{algorithm}

\rev{\emph{On the nature of the bound.} RMSC is modular: its error
terms (truncation, Trotter, statistical, plus the depth-deterministic gate
noise) are decoupled, and each module admits independent alternatives
and corresponding optimization algorithms. Theorem~\ref{thm:perf} therefore
states an \emph{upper bound}: for a fixed circuit-depth budget, a suitable
combination of modules keeps the sample count within the stated order of
magnitude, and different combinations trace out a trade-off frontier. For
example, a double Monte-Carlo sampling scheme can reduce the per-sample
circuit depth to $O(1)$ at the cost of a sharp increase in the sample count;
conversely, a spectral-simulation module based on QSVT can drastically reduce
$N$ but requires longer circuits or full knowledge of the state-generation
black box. Theorem~\ref{thm:perf} describes one favorable operating point of
this frontier, chosen for current hardware levels. The
statistical deviation of the estimator from its regularized
target $E^{(\eta)}_t$ is controlled by the same theorem. The
regularization itself leaves a deterministic,
mechanism-dependent distance between $E^{(\eta)}_t$ and the true
divergence $E_t$; the following theorem gives this distance
exactly.}

\begin{theorem}[regularization gap]\label{thm:main}
\rev{Let $t\ge1$ be fixed (constants may depend at most polynomially on
$t$), let $\eta\in(0,1)$ be the regularization level, and denote the
\emph{certification gap} by $G_\eta(\rho\|\sigma)$.
Then, for every pair of states $\rho$ and $\sigma$,
\begin{align}\label{eq:sandwich}
E_t(\rho\|\sigma)&=E^{(\eta)}_t(\rho\|\sigma)+G_\eta(\rho\|\sigma),\\
G_\eta(\rho\|\sigma)&:=\frac12\sum_k|\lambda_k|\operatorname{erfc}
\!\left(\frac{|\lambda_k|}{2\sqrt{\eta}}\right).\label{eq:Geta_def}
\end{align}}
\end{theorem}

\rev{The proof is given in Appendix~\ref{app:thm1}. Since every term of $G_\eta$ is
nonnegative, the regularized target never exceeds the truth,
$E^{(\eta)}_t\le E_t$, with equality if and only if $G_\eta=0$. For
every $c>0$ the gap obeys the split bound
\begin{equation}\label{eq:split}
G_\eta\le\tfrac12\,W_{c\sqrt\eta}+\tfrac12(1+t)e^{-c^{2}/4},
\end{equation}
where $W_{c\sqrt\eta}$ is the total spectral weight of $H$ within
$\pm c\sqrt\eta$ of zero: at most half the near-null spectral mass
plus an exponentially small out-of-window tail. The split bound is proved together with Theorem~\ref{thm:main}
in Appendix~\ref{app:thm1}. $E^{(\eta)}_t$
increases monotonically to $E_t$ as $\eta\to0$; the regularization
is thus a controlled approximation, not a distortion.}

\rev{\emph{Success probability and statistical fluctuation.} By
Theorem~\ref{thm:perf}, a single audit run succeeds with probability at
least $1-\beta$; on the failure runs the estimate may deviate in either
direction, so any lower bound on $E_t$ must guard against the
\emph{upward} fluctuation of $\hat E_t$ and subtract from it a
\emph{statistical allowance} $s_\beta$. The allowance is obtained
by inverting the one-sided Bernstein inequality for the Monte-Carlo
samples; its definition from the per-sample variance and range of
the estimator is given in Appendix~\ref{app:thm3}. All statements of the
compensation below hold on the anti-jump event
\begin{equation}\label{eq:Abeta}
\mathcal A_\beta:=\bigl\{\hat E_t\le E^{(\eta)}_t+s_\beta\bigr\},
\qquad \Pr[\mathcal A_\beta]\ge1-\beta .
\end{equation}
The regularization gap $G_\eta$ always pushes the expectation of
$\hat E_t$ below the truth. A natural question follows: how
much of this deterministic gap can be paid back by a purely
classical post-processing of the already-measured $\hat E_t$, and
at what confidence? Theorem~\ref{thm:inv} and
Corollaries~\ref{cor:cert} and~\ref{cor:rec} answer this
question.}

\begin{lemma}[structural floor]\label{lem:floor}
\rev{Let $w:=2\sqrt\eta$ be the spectral-resolution window and define
$\varphi(E):=\tfrac12 E\operatorname{erfc}(E/w)$. Every state pair
satisfies
\begin{equation}\label{eq:floor}
G_\eta\;\ge\;\varphi(E_t).
\end{equation}
Moreover, $\varphi$ is strictly increasing on $[0,E^{*}]$ and strictly
decreasing on $[E^{*},\infty)$, where $E^{*}=w u^{*}\approx1.06\sqrt\eta$
and $u^{*}\approx 0.532$ is the unique positive root of
$\operatorname{erfc}(u)=\frac{2}{\sqrt\pi}\,u\,e^{-u^{2}}$; its maximum
is $\varphi_{\max}=\varphi(E^{*})\approx0.240\sqrt\eta$.}
\end{lemma}

\rev{The proof is given in Appendix~\ref{app:thm3}. Combined with
Theorem~\ref{thm:main}, Lemma~\ref{lem:floor} implies that the
regularized target satisfies $E^{(\eta)}_t = E_t - G_\eta \le
E_t - \varphi(E_t)$ for every pair of states. This universal upper
envelope motivates defining a forward bias map $\mathrm{F}(E):=E-\varphi(E)$
and applying its inverse $\mathrm{F}^{-1}$ directly to the measured estimate
$\hat E_t$ to compensate for the regularization gap. Because $\mathrm{F}^{-1}$
operates solely on the scalar output $\hat E_t$ via a one-dimensional
monotone root-finding routine (evaluated in microseconds on a classical
processor), this compensation requires no additional quantum samples,
circuits, or oracle calls.}

\begin{theorem}[faithful inversion]\label{thm:inv}
\rev{Let $\mathrm{F}(E):=E-\varphi(E)=E-\tfrac{E}{2}\operatorname{erfc}(E/w)$ on
$[0,\infty)$, with $\mathrm{F}^{-1}$ denoting its inverse (extended by the
identity map on $(-\infty,0)$), and define the compensated point estimator
$\hat E_{\rm inv}:=\mathrm{F}^{-1}(\hat E_t)$. Then, for every state pair,
\begin{equation}\label{eq:thminv}
\mathrm{F}^{-1}\bigl(\E[\hat E_t]\bigr)\le E_t,
\qquad
\E[\hat E_{\rm inv}]\ge\E[\hat E_t].
\end{equation}}
\end{theorem}

\rev{The proof is given in Appendix~\ref{app:thm3}. The map $\mathrm{F}$ is the exact forward bias model of
RMSC in the single-dominant-eigenvalue regime: if the positive
part of the spectrum of $H=\rho-t\sigma$ consists of one eigenvalue
$\lambda_+=E_t$ with all remaining eigenvalues far from the window,
then $G_\eta=\varphi(E_t)$ and $\E[\hat E_t]=\mathrm{F}(E_t)$. The map $\mathrm{F}$ is
a strictly increasing bijection of $[0,\infty)$ onto itself,
$\mathrm{F}'(E)=1-\varphi'(E)\ge\tfrac12>0$, so $\mathrm{F}^{-1}$ exists and is
computable as a one-dimensional monotone root-find, in microseconds on
a classical processor. Two properties make the inversion rigorous far
beyond this regime: the unconditional floor $G_\eta\ge\varphi(E_t)$
of Lemma~\ref{lem:floor}, i.e., $\E[\hat E_t]\le \mathrm{F}(E_t)$ for every
state pair; and the monotonicity of $\mathrm{F}^{-1}$, which transports
one-sided inequalities unchanged and is the reason a statistical
allowance subtracted \emph{before} the inversion is inherited for
free afterwards. The equality case of the first bound requires all
positive eigenvalues to coincide with $E_t$: for spectra with a single
dominant positive eigenvalue the estimator is therefore asymptotically
unbiased, $\E[\hat E_{\rm inv}]\to E_t$ as $N\to\infty$, while the
second bound says the compensation never degrades the mean accuracy.}

\rev{The inversion is a deterministic map and creates no randomness of
its own; the two variance notions at play serve different roles. The
per-sample variance $V$ and the allowance $s_\beta$ live in the input
space: they are fixed by $(N,t,Z,\beta)$ before any post-processing and
alone determine the confidence level of every statement above. The
variance of $\hat E_{\rm inv}$, by contrast, describes how the same
sampling noise reappears in the compensated output: wherever
the statistical noise of $\hat E_t$ is small compared with
$\mathrm{F}(E_t)$,
\begin{equation}\label{eq:vartrans}
\Var[\hat E_{\rm inv}]
=\bigl(1-\varphi'(E_t)\bigr)^{-2}\Var[\hat E_t]\,\bigl(1+o(1)\bigr),
\end{equation}
with factor $K=(1-\varphi')^{-2}$ obeying $K(0)=4$ and $K<1$ on
$(E^{*},\sim\!2.5w)$ (Appendix~\ref{app:thm3}). It therefore determines
the single-run spread of the point estimate, that is, its mean-squared
accuracy as an estimator. It enters nothing else: neither $s_\beta$ nor any
confidence statement. Near the origin the compensated estimate may thus
be noisier than the raw one in single-run mean-squared error (the
bias--variance price of removing the gap; recoverable, if desired, by a
one-parameter shrinkage
$\lambda\hat E_{\rm inv}+(1-\lambda)\hat E_t$), while on the decreasing
branch of $\varphi$ the inversion even de-amplifies noise. For spectra
with many eigenvalues inside the window the residual bias
$-(G_\eta-\varphi(E_t))$ is unavoidable classically (it depends on the
spectrum, not on $E_t$), but keeps the one-sided sign: the compensated
estimate never overestimates in expectation.}

\begin{table*}[t]
\centering
\caption{Comparison of RMSC with prior approaches
for estimating the quantum Hockey--stick divergence from copies of the
states. Column meanings: \emph{Access}---how each method is allowed
to interact with the states (consuming copies only, or requiring an
additional oracle); \emph{Sample complexity}---number of samples needed to
reach precision $\varepsilon$; \emph{Depth}---circuit depth per sample and
its scaling with the system size $n$; \emph{Main limitation}---the principal
cost or restriction of each method. All methods operate on $O(n)$-qubit
registers; ``$n$-free'' means independent of $n$.}
\label{tab:comparison}
\small
\begin{tabular}{|l|c|c|c|c|}
\hline
Method & Access & Sample complexity & Depth & Main limitation \\
\hline
Ours (RMSC) & copies only & $\widetilde{O}(\varepsilon^{-2}\ln^2\frac1\varepsilon)$, $n$-free & $O(\varepsilon^{-2})$, $n$-free & requires $n$ fresh copies per Trotter step \\
\hline
QST & copies only & $O(2^{2n})$--$O(2^{3n})$ & $O(1)$ & exponential samples \\
\hline
CST & copies only & $O(2^{n})$--$O(2^{2n})$ & $O(1)$ & exponential for nonlinear $E_t$ \\
\hline
QSVT & block-encoding oracle & $\mathrm{poly}(1/\varepsilon)$ & $\mathrm{poly}(1/\varepsilon)$ & oracle unavailable in auditing \\
\hline
IQPE$+$LMR & copies only & $O(n^8)$ time & exponential & high time; exponential depth \\
\hline
VQA & copies only & heuristic & shallow & no rigorous guarantee \\
\hline
\end{tabular}
\end{table*}

\begin{corollary}[safe lower bound]\label{cor:cert}
\rev{Let $\mathrm{L}_0=\hat E_t-s_\beta$ denote the uncompensated baseline
lower bound on the event $\mathcal A_\beta$ of Eq.~(\ref{eq:Abeta}).
Then, on $\mathcal A_\beta$, every state pair satisfies
\begin{equation}\label{eq:linv}
E_t\;\ge\;\mathrm{L}_{\rm inv}:=\mathrm{F}^{-1}\bigl(\hat E_t-s_\beta\bigr),
\end{equation}
requiring no statistical allowance beyond the quantity $s_\beta$ already
subtracted in $\mathrm{L}_0$.}
\end{corollary}

\rev{The proof is given in Appendix~\ref{app:thm3}. The confidence
guarantee holds in the input space: writing $x=\hat E_t-s_\beta$,
the event $\mathcal A_\beta$ and Lemma~\ref{lem:floor} ensure that
$x\le E^{(\eta)}_t=E_t-G_\eta\le \mathrm{F}(E_t)$, and the strictly increasing
map $\mathrm{F}^{-1}$ transports this one-sided inequality directly. Consequently,
the variance amplification analyzed in Theorem~\ref{thm:inv} does not
erode the validity of the lower bound: the confidence level is inherited
from $\mathcal A_\beta$ by monotonicity rather than re-derived in the
output space, so the same pair $(N,s_\beta)$ that certifies $\mathrm{L}_0$
also certifies $\mathrm{L}_{\rm inv}$.}

\rev{Before comparing with simpler corrections, we recall why the
direct $\varphi$ compensation is not adopted as the final construction:
the one-step additive correction $x\mapsto x+\varphi(x)$ is the
first-order expansion of the inversion. It is valid as a lower bound only
when the (unknown) truth lies below the floor peak, $E_t\le E^{*}$, and
retains a composition bias of order $\varphi\,\varphi'$. With this in
mind, the relation between the two is simple: $\mathrm{L}_{\rm inv}$ is the
fixed point of the iteration $y\mapsto x+\varphi(y)$ started at
$y_0=x$, whose first iterate is precisely the one-step compensation;
$\mathrm{L}_{\rm inv}$ thus resums all higher orders of that construction at
once and dominates it on $0\le x\le \mathrm{F}(E^{*})$
(Appendix~\ref{app:thm3}). For single-dominant-eigenvalue spectra
$\mathrm{L}_{\rm inv}\to E_t$ as $N\to\infty$: no composition bias. Unlike the
one-step bound, the corollary is range-unconditional. This is essential in
black-box auditing, where no side condition on $E_t$ can be checked. If
the mechanism family is white-box known so that $G_\eta$ is computable,
the ideal bound $\mathrm{L}_2:=\hat E_t+G_\eta-s_\beta$ of the earlier
formulation remains the reference construction.}

\begin{corollary}[recovery interval]\label{cor:rec}
\rev{Let $x_L$ and $x_R$ denote the two roots of
$\varphi(x)=s_\beta$, ordered so that $x_L<x_R$. Then
\begin{equation}\label{eq:recovery}
\E[\mathrm{L}_{\rm inv}]>\E[\hat E_t]
\end{equation}
if and only if the compensated lower bound lies between the two roots,
$\mathrm{L}_{\rm inv}\in(x_L,\,x_R)$. These roots exist if and only if
$s_\beta<0.240\sqrt\eta$, and admit the closed-form approximations
\begin{equation}\label{eq:endpoints}
\begin{aligned}
x_L&\approx\tfrac{w\sqrt\pi}{4}\Bigl(1-\sqrt{1-\tfrac{16\,s_\beta}
{w\sqrt\pi}}\Bigr)\approx2\,s_\beta,\\
x_R&\approx w\,\sqrt{\ln\tfrac{w}{2\sqrt\pi\,s_\beta}}\,,
\end{aligned}
\end{equation}}
\end{corollary}

\rev{The proof is given in Appendix~\ref{app:thm3}. Since $\mathrm{F}^{-1}$ is increasing, the activation condition
(\ref{eq:recovery}) reads, in raw units,
$\E[\hat E_t]\in(x_L,\,x_R)$, and, in truth units for
single-dominant spectra,
$E_t\in\bigl(\mathrm{F}^{-1}(x_L),\,\mathrm{F}^{-1}(x_R)\bigr)$. On this interval the
inversion wins back the confidence allowance, so recovery replaces
loss. Outside it the safe lower bound remains valid but no
longer beats the raw estimate, its role there being confidence rather
than recovery. The proof and the performance metrics of the interval
(accuracy of (\ref{eq:endpoints}), existence threshold in $N$, and the
matched-window design) are given in Appendix~\ref{app:thm3}.}

\section{Comparison and performance}\label{sec:perf}

\rev{This section positions RMSC among existing estimators of
the Hockey--stick divergence and assesses its near-term
implementability.}

\emph{\rev{Comparison.}} \rev{Table~\ref{tab:comparison} summarizes
the competing routes in the copies-only setting; the families are
introduced in the Introduction and the caption, so we comment only on
the axis that matters here. Every alternative pays for the divergence
with an exponential resource: copies (QST, CST), circuit depth
(IQPE), or an oracle unavailable in auditing (QSVT). Alternatively it forgoes
rigorous guarantees altogether (VQA). RMSC is
distinctive in that \emph{both} of its principal resources are $n$-free
and polynomial: the sample complexity
$\widetilde O(\varepsilon^{-2}\ln^{2}(1/\varepsilon))$ of
Theorem~\ref{thm:perf} and the typical per-sample depth
$O(\varepsilon^{-2})$ gate layers. It thus occupies a favorable point
of the depth--sample balance---shallow circuits acting on fresh copies,
repeated $N$ times, with circuit width $O(n)$---and the balance is
tunable: the depth reductions (trace-level Trotter, Richardson
extrapolation) and the depth--sample frontier are documented in
Appendix~\ref{app:opt}.}

\emph{\rev{NISQ feasibility.}}\label{sec:nisq}
\rev{The more demanding question is whether the required depth,
fidelity, and sampling volume fit on existing hardware. Frontier
superconducting and trapped-ion processors report two-qubit gate error
rates of $1.8\times10^{-3}$ to $3.3\times10^{-3}$ (Google Willow CZ
error $0.33\%$ \cite{acharya25}; Quantinuum H2 two-qubit infidelity
$1.84\times10^{-3}$ \cite{moses23}; IBM Heron-class median CZ error
$\approx3\times10^{-3}$ \cite{sobri25}), entangling-gate durations of
$\approx30$--$90$~ns, and $T_{1}\approx68$--$98~\mu$s coherence times
\cite{acharya25}.}
\rev{Consider, as a concrete task, the estimation of
$E_{e^{\epsilon_{\mathrm{DP}}}}(\rho\|\sigma)$ for a single-qubit pair
of states---a single-qubit audit task of the same class as those of Section~\ref{sec:experiments}. If
every hardware metric is pushed to the frontier levels above---per-gate
error $\gamma\approx2\times10^{-3}$ and per-sample circuits of
$\approx10^{2}$ gate layers on average ($\lesssim4\times10^{2}$ at the
deepest; Table~\ref{tab:optres}), each executing within a few $\mu$s, several times below the
coherence budget---the verified resource counts
(Table~\ref{tab:optres}; Appendix~\ref{app:resnoise}) are, to order of
magnitude: a precision $\varepsilon\approx0.1$ is reached with
$N\approx3\times10^{3}$ samples per side, i.e.\ $10^{3}$ samples and
$10^{5}$--$10^{6}$ total state copies, with the end-to-end error below
$\varepsilon$ in every run; tightening to $\varepsilon\approx0.05$
raises the counts to $\sim2\times10^{4}$ samples per side and
$\sim10^{7}$ copies, consistent with the $\varepsilon^{-2}$ law of
Theorem~\ref{thm:perf}. The noise-bias budget requires
$\gamma\lesssim\varepsilon^{3}$ up to logarithms
(Appendix~\ref{app:resnoise}); the $\varepsilon\approx0.1$ window
$\gamma\lesssim1.7\times10^{-3}$ is within a factor of $\approx2$ of
all three platforms above, and the residual noise bias, being a
deterministic function of the depth, is further suppressible by
zero-noise extrapolation \cite{sobri25}. At the precision where the
depth--fidelity product saturates frontier hardware, RMSC
therefore delivers a complete divergence estimate---together with the
compensated point estimate and the safe lower bound of
Section~\ref{sec:algorithm}---from $\sim\!10^{5}$ copies, which to our
knowledge makes it the first divergence estimator operable at practical
precision on present-day hardware. The end-to-end case study is presented in
Section~\ref{sec:experiments}.}

\section{Numerical experiments}\label{sec:experiments}

This section validates the theoretical guarantees of the preceding
sections---the sample-complexity law (Theorem~\ref{thm:perf}), the
regularization gap (Theorem~\ref{thm:main}), the inversion
compensation (Theorem~\ref{thm:inv}), and the end-to-end estimation of
the divergence for quantum differential privacy (QDP)
mechanisms---on the full RMSC pipeline: every
data point is a genuine Monte-Carlo run with importance sampling
(inverse-CDF of the $E_1$-type density), single-shot $\pm1$ Hadamard
readouts, and the analytic small-region correction; oracle quantities
($E_t$, $E^{(\eta)}_t$, $G_\eta$) are computed once by exact
diagonalization of $H=\rho-t\sigma$ and are used only for validation.

\begin{figure*}[t]
\centering
\includegraphics[width=\textwidth]{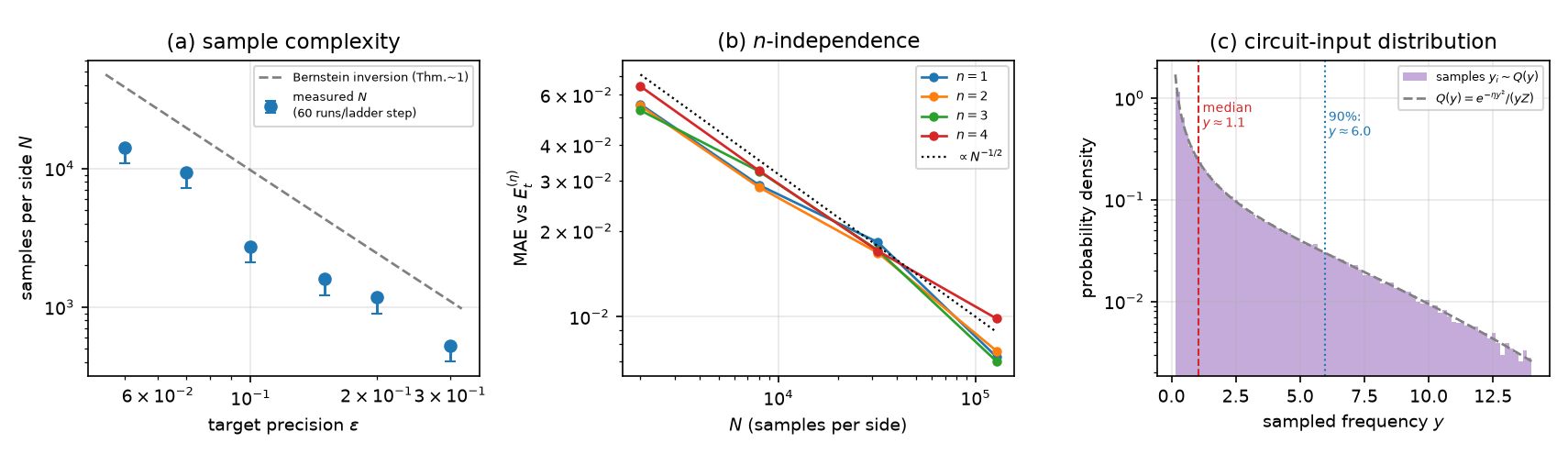}
\caption{Sample complexity. (a) Measured budget $\widehat N$ vs.\
target $\varepsilon$: points are the first ladder rung at which
$60$-run empirical success reaches $1-\beta$ (bracketed from below
by the failing rung); the dashed line is the combined-estimator Bernstein prescription (Appendix~\ref{app:thm2}). (b) Mean absolute error vs.\
$E^{(\eta)}_t$ for $n=1,\dots,4$ qubits. (c) Distribution of the
sampled circuit-input frequencies $y_i$ (histogram) with the
importance density $Q(y)$ (dashed); the marked percentiles bound the
per-sample circuit depth, which is proportional to $y$.}
\label{fig:E1}
\end{figure*}

\emph{Common protocol.} Unless stated otherwise, all experiments adopt
the audit parameters $t=e^{\epsilon_{\mathrm{DP}}}=e$
($\epsilon_{\mathrm{DP}}=1$), privacy threshold $\delta=0.1$,
regularization parameter $\eta=0.01$, failure probability $\beta=0.05$, and
sampling cutoff $\alpha=0.15$. Under these settings, Eq.~(\ref{eq:Q}) determines
the normalization constant $Z=\tfrac12E_1(\eta\alpha^{2})=3.91$;
Lemma~\ref{lem:isvar} gives the per-sample variance bound
$V=(1+t^{2})(Z/\pi)^{2}=13.0$ and range $R=(1+t)(Z/\pi)=4.63$; and
Theorem~\ref{thm:perf} defines the statistical allowance $s_\beta$ as the positive
root of $Ns^{2}=(2V+\tfrac23Rs)\ln(1/\beta)$, yielding $s_\beta=0.0114$
at the reference budget of $N=6\times10^{5}$ samples per side. The
spectral-resolution window $w=2\sqrt{\eta}$, the bias map $\mathrm{F}$, and its inverse $\mathrm{F}^{-1}$
follow Lemma~\ref{lem:floor}, Theorem~\ref{thm:inv}, and
Corollary~\ref{cor:cert} in Section~\ref{sec:algorithm}. The test states are $n$-qubit state pairs transmitted
through a depolarizing channel $\mathcal D_p(X)=(1-p)X+p\,I/2^{n}$;
each experiment specifies its state pairs and, where applicable, its values of
$\eta$ and $\alpha$. Circuit-depth evaluations use a Trotter step size $\Delta y=0.11$ and count gate layers as $2M(y)$, where $M(y)=\lceil y/\Delta y\rceil$.

\subsection{Sample complexity of the estimator}

\emph{Setup.} This experiment evaluates the sample-complexity scaling of
Theorem~\ref{thm:perf},
$N=\widetilde O(\varepsilon^{-2}\ln^{2}(1/\varepsilon))$ at confidence level
$1-\beta$, examining three aspects: the growth of the required sample size $N$ as the target precision
$\varepsilon$ is tightened, the independence of the estimation error from the system
size $n$ at fixed $N$, and the resulting circuit-depth distribution on the quantum
processor. All three panels use the default values of $\eta$ and $\alpha$ from
the common protocol on depolarizing state pairs.

In panel (a), for each target precision
$\varepsilon\in\{0.05,0.07,0.1,0.15,0.2,0.3\}$, the required sample
count is determined empirically: starting from
$N_0=\tfrac16 N_{\mathrm{B}}$, where
$N_{\mathrm{B}}=(2V+\tfrac23R\varepsilon)\ln(2/\beta)/\varepsilon^{2}$ is the combined-estimator Bernstein bound,
the sample budget is increased geometrically by a factor of $1.3$ per step until the
empirical success rate
$\widehat P(|\hat E_t-E^{(\eta)}_t|\le\varepsilon)$, evaluated over
$60$ independent pipeline runs at each step, reaches the target confidence
$1-\beta$. The recorded sample complexity $\widehat N$ is the smallest budget meeting this threshold, with the immediately preceding budget shown as a lower bracket.

In panel (b), the system size is varied from $n=1$ to $n=4$ qubits across four sample
budgets $N\in[2\times10^{3},1.28\times10^{5}]$ ($40$ runs per data point).

Panel (c) shows the empirical distribution of the sampled evolution times $y_i$ passed to the
quantum circuits ($2\times10^{5}$ draws), where $M(y)=\lceil y/\Delta y\rceil$ directly determines the per-sample circuit depth.

\emph{Results.} The results are presented in Fig.~\ref{fig:E1}.

In panel (a), the empirically required
sample counts $\widehat N$ follow the $\varepsilon^{-2}$ scaling across the sixfold range of $\varepsilon$ (appearing nearly linear on log--log axes,
since for $\varepsilon\le0.3$ the range correction $\tfrac23R\varepsilon$
alters the pure power law by less than $6\%$) and lie consistently below
the worst-case Bernstein bound of Theorem~\ref{thm:perf} by a factor of
$2$ to $3.5$, confirming that the theoretical bound is safely conservative while the achieved precision reliably satisfies the target.

In panel (b), at each fixed $N$, the mean absolute error curves for
$n=1,\dots,4$ qubits coincide within statistical fluctuations and decay
as $N^{-1/2}$, demonstrating that the sample complexity is independent of the Hilbert-space
dimension and retains the canonical $N^{-1/2}$ Monte-Carlo convergence rate.

In panel (c), the sampled frequencies concentrate at small values of $y$ (with a median of
$y\approx1.1$ and a $90$th percentile of $y\approx6.0$), whereas the importance-sampling
density avoids the small-$y$ regime $y<\alpha$ where shot
noise would otherwise diverge. Because the per-sample circuit depth scales linearly
with $y$, these percentiles correspond directly to hardware circuit depths: $50\%$ of all samples require at most $\approx20$ gate layers ($10$ Trotter factor pairs) and
$90\%$ require fewer than $\approx110$ layers, ensuring that typical executions remain shallow
and deep circuits occur with exponentially small probability.

\subsection{The regularization gap}

\emph{Setup.} This experiment verifies the exact decomposition of
Theorem~\ref{thm:main}, $E_t=E^{(\eta)}_t+G_\eta$ with
$G_\eta=\frac12\sum_k|\lambda_k|\operatorname{erfc}(|\lambda_k|/w)$,
across two control parameters: varying the positive eigenvalue $\lambda_+$ at a fixed window width $w$,
and varying the window width $w$ for a fixed state pair. Because the quantity of interest is the
difference $E_t-\overline{\hat E_t}$ between the exact divergence and
the mean estimator output, the small-region truncation bias must be kept negligible relative to $G_\eta$; both sweeps therefore use a
smaller cutoff $\alpha=0.08$ (truncation bias $\le6\times10^{-4}$), while keeping all
other parameters at their common-protocol values with $N=6\times10^{5}$
samples per side and $16$ independent runs per data point.

Panel (a) fixes $w=0.1$ ($\eta=2.5\times10^{-3}$) and varies the
single positive eigenvalue $\lambda_+\in[0.02,0.56]$ by adjusting the depolarizing noise strength.

Panel (b) fixes the state pair at $E_t=0.1$ and varies $w\in[0.05,1.0]$, spanning four orders of magnitude in $\eta$.

\begin{figure*}[t]
\centering
\includegraphics[width=\textwidth]{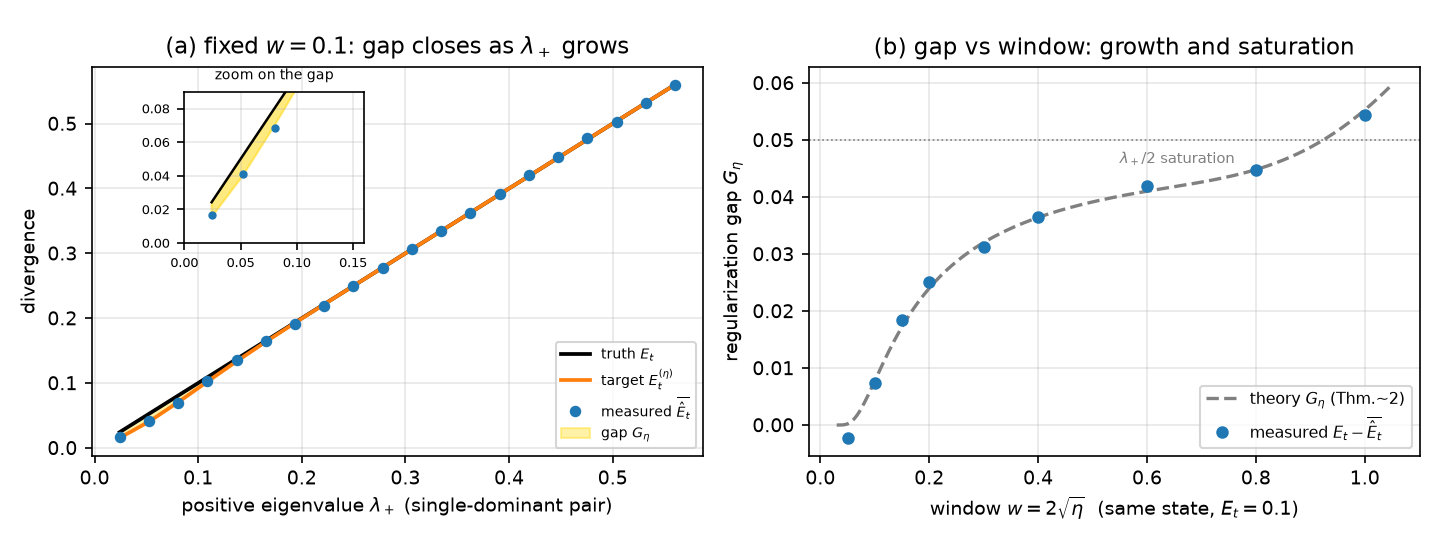}
\caption{The regularization gap. (a) Fixed $w=0.1$: truth $E_t$,
target $E^{(\eta)}_t$, and measured estimates vs.\ the positive
eigenvalue; the gold band is the gap $G_\eta$ (inset: zoom on small
$\lambda_+$). (b) Fixed state ($E_t=0.1$): theory $G_\eta$ of
Theorem~\ref{thm:main} (dashed) and measured gap (points) vs.\ the
window $w$. $\alpha=0.08$, $N=6\times10^{5}$, $16$ runs per point.}
\label{fig:E2}
\end{figure*}

\emph{Results.} Figure~\ref{fig:E2} displays the results of both sweeps.

In panel (a), the mean estimates closely match
the regularized target $E^{(\eta)}_t$ across all values of $\lambda_+$. Consequently, the
regularization gap $G_\eta$ (gold shaded region) emerges for small $\lambda_+$, attains a maximum near
$\lambda_+\approx w/2$, and vanishes once $\lambda_+$ substantially exceeds $w$: spectral components with $\lambda\lesssim w$ lose approximately half of their
magnitude to regularization, whereas components with $\lambda\gg w$ are preserved almost intact.

In panel (b), for a fixed state pair, the empirical gap agrees with the theoretical curve $G_\eta$ across the entire range of $w$ and saturates at $\lambda_+/2$ for $w\gg\lambda_+$, corresponding to the asymptotic regime in which the smooth ramp reduces the positive eigenvalue by half. These results confirm Theorem~\ref{thm:main} in both regimes and delineate the operating boundary of the uncompensated estimator: when $w$ is much smaller than the relevant spectral scale, the regularization bias is negligible, whereas when $w$ is comparable to the eigenvalues, the resulting gap $G_\eta$ must be removed via classical inversion, as demonstrated next.

\subsection{Inversion compensation across the operating range}

\emph{Setup.} This experiment evaluates the classical inversion compensation of
Theorem~\ref{thm:inv}---comprising both the compensated point estimate
$\hat E_{\rm inv}=\mathrm{F}^{-1}(\hat E_t)$ and the certified lower bound
$\mathrm{L}_{\rm inv}=\mathrm{F}^{-1}(\hat E_t-s_\beta)$ of Corollary~\ref{cor:cert}---relative to the
uncompensated estimator $\hat E_t$ across the full parameter range of the depolarizing channel. We
vary the depolarizing probability $p\in[0,0.412]$ over $109$ steps on the
reference state pair, so that $E_t$ decreases continuously from the small-gap regime ($G_\eta\to0$ as $p\to0$), through the
audit threshold ($E_t=\delta$ at $p=0.337$) and the theoretical recovery
interval, down to $E_t=0$. Each data point represents the mean over $24$ independent
full-pipeline runs with $N=6\times10^{5}$ samples per side.

\begin{figure*}[t]
\centering
\includegraphics[width=\textwidth]{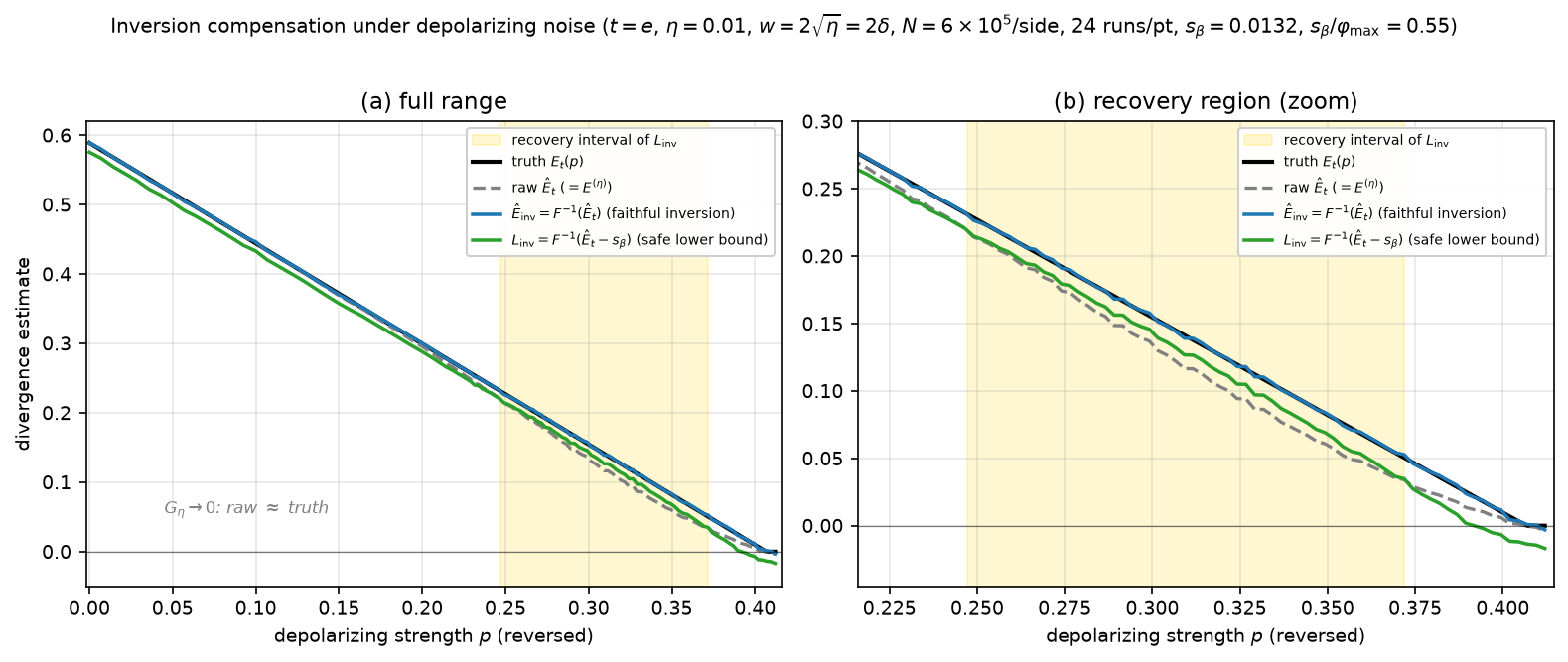}
\caption{Inversion compensation under depolarizing noise. (a) Full
range $p\in[0,0.412]$; (b) zoom on the recovery region. Black:
truth; gray dashed: raw estimate; blue:
$\hat E_{\rm inv}=\mathrm{F}^{-1}(\hat E_t)$; green:
$\mathrm{L}_{\rm inv}=\mathrm{F}^{-1}(\hat E_t-s_\beta)$. Gold band: recovery interval
of Corollary~\ref{cor:rec}. $24$ runs per point, $N=6\times10^{5}$
per side.}
\label{fig:E3}
\end{figure*}

\emph{Results.} Figure~\ref{fig:E3} summarizes the results.

First, the compensated point estimate $\hat E_{\rm inv}$
accurately tracks the true divergence $E_t$ across the entire parameter sweep: at the audit threshold
($E_t=0.101$, $G_\eta=0.024$), it yields a mean value of $0.1009$, recovering $98\%$ of
the regularization gap, and as $p\to0$ (right boundary) it converges to the uncompensated estimate as
$\mathrm{F}^{-1}$ approaches the identity map.

Second, the certified lower bound $\mathrm{L}_{\rm inv}$
exceeds $E_t$ in only $0.7\%$ of all $2616$ realizations (well within the nominal failure rate $\beta=5\%$), and
its sample mean surpasses the uncompensated mean $\overline{\hat E_t}$ precisely within the recovery interval predicted by
Corollary~\ref{cor:rec}: the roots $x_L=0.0323$ and $x_R=0.2196$ of
$\varphi(x)=s_\beta$ correspond to true divergences $E_t\in(0.051,0.231)$, or $p\in(0.247,0.372)$ (gold shaded band), inside
which the green curve lies above the uncompensated curve and outside which it lies below.

Third, the empirical standard-deviation ratio
$\mathrm{std}(\hat E_{\rm inv})/\mathrm{std}(\hat E_t)$ remains within $[0.89,1.63]$ across the entire sweep (consistent with the theoretical variance-transfer factor $K\in[0.785,4]$ in
Appendix~\ref{app:thm3})
and stays within $[0.90,1.05]$ inside the recovery
interval, confirming that the inversion introduces negligible statistical overhead near the audit threshold.

\subsection{Estimating the divergence of QDP mechanisms}

\emph{Setup.} This experiment demonstrates the end-to-end application of RMSC to the black-box
auditing of quantum differential privacy (QDP). A quantum channel $\mathcal C$ is
$(\epsilon_{\mathrm{DP}},\delta)$-differentially private if and only if
$E_{e^{\epsilon_{\mathrm{DP}}}}(\mathcal C(\rho)\|\mathcal C(\sigma))\le\delta$
for all quantum-adjacent input pairs satisfying $\frac12\|\rho-\sigma\|_1\le1$
\cite{hrf23,zy17,ar19}; in a black-box audit, the auditor accesses the channel outputs solely through independent state
copies \cite{gfy23,adk23}, matching the copies-only setting introduced in Section~\ref{sec:intro}. We evaluate three standard single-qubit noise channels---depolarizing, erasure, and dephasing (whose definitions, parameters, and
references are detailed in Appendix~\ref{app:qdp})---on input pairs prepared at three
trace distances $d\in\{0.3,0.6,1.0\}$, where $d=1.0$ corresponds to the QDP
adjacency boundary. The channel noise parameters are chosen so that the three input distances
probe distinct regimes relative to the privacy threshold $\delta=0.1$: moderately distinguishable inputs ($d=0.6$) yield output divergences near $\delta$, boundary inputs ($d=1.0$) produce divergences substantially above $\delta$ ($E_t=0.54$, $0.75$, and $1.0$), and weakly distinguishable inputs ($d=0.3$) have their output divergences contracted to zero. For each of the nine configurations, the full RMSC pipeline
(using the common protocol with $N=3\times10^{5}$ samples per side averaged over $24$ runs) produces the raw estimate, applies the faithful inversion $\mathrm{F}^{-1}$ projected onto
$[0,\infty)$, and compares the result against the exact divergence computed numerically from the $2\times2$ output density matrices.

\emph{Results.} Table~\ref{tab:audit} summarizes the nine audit configurations, where \emph{err per} denotes the ratio of the mean estimate to the exact value expressed as a percentage ($100\%$ indicating zero bias). Across all nine entries, the compensated estimates agree closely with the exact values, exhibiting absolute errors between $0$ and
$0.0035$---comparable to the multi-run statistical standard error $\sqrt{V/(24N)}\approx1.3\times10^{-3}$---and relative ratios within $[99.4\%,101.8\%]$. For all three channels at $d=0.3$, data processing reduces the true output divergence to zero, and the estimator
accurately returns $0.0000$ without false positives. At the adjacency boundary
($d=1.0$), which determines the worst-case privacy loss, the true divergences $E_t=0.5352$,
$0.7500$, and $1.0000$ are resolved with relative errors between $0.1\%$ and $0.35\%$, unambiguously certifying the margin by which each mechanism exceeds the threshold $\delta=0.1$. Moreover, all nine audits execute on circuits whose resource demands fall well within the NISQ feasibility envelope established in Section~\ref{sec:perf} (requiring a mean circuit depth of
$\approx41$ gate layers, a $99$th-percentile depth below $\approx235$ layers, and compatible per-gate error tolerances of $\gamma\approx1.7\times10^{-3}$ to
$1.3\times10^{-4}$).

\begin{table}[htbp]
\centering
\caption{Divergence estimation for three QDP channels at three input
trace distances $d$ ($N=3\times10^{5}$ samples per side, $24$ runs).}
\label{tab:audit}
\begin{tabular}{lccccc}
\hline
Mechanism & $d$ & $E_t$ (truth) & estimate & $|$error$|$ & err per \\
\hline
Depolarizing   & 0.3 & 0.0000 & 0.0000 & 0.0000 & 100\% \\
$p=0.25$       & 0.6 & 0.1235 & 0.1241 & 0.0006 & 100.5\% \\
               & 1.0 & 0.5352 & 0.5360 & 0.0008 & 100.1\% \\
\hline
Erasure        & 0.3 & 0.0000 & 0.0000 & 0.0000 & 100\% \\
$p_{\mathrm{er}}=0.25$ & 0.6 & 0.1657 & 0.1687 & 0.0030 & 101.8\% \\
               & 1.0 & 0.7500 & 0.7511 & 0.0011 & 100.1\% \\
\hline
Dephasing      & 0.3 & 0.0000 & 0.0000 & 0.0000 & 100\% \\
$p=0.15$       & 0.6 & 0.0620 & 0.0616 & 0.0004 & 99.4\% \\
               & 1.0 & 1.0000 & 1.0035 & 0.0035 & 100.3\% \\
\hline
\end{tabular}
\end{table}

\section{Conclusion}\label{sec:conclusion}

In this work, we have addressed the problem of estimating the quantum Hockey--stick divergence $E_t(\rho\|\sigma)$ solely from copies of unknown quantum states on near-term NISQ hardware, a task previously hindered by exponential overheads, inaccessible oracles, or a lack of rigorous guarantees. To this end, we proposed RMSC and established three main results. First, RMSC achieves a qubit-independent sample complexity of $\widetilde{O}(\varepsilon^{-2}\ln^{2}(1/\varepsilon))$ with a shallow per-sample circuit depth of $O(\varepsilon^{-2})$, matching current frontier hardware capabilities at precision $\varepsilon\approx 0.1$. Second, we exactly characterized the deterministic regularization gap $G_\eta$ and eliminated it via a purely classical analytic inversion of the bias map; we further constructed a certified safe lower-bound compensation scheme and derived closed-form expressions for its effective recovery interval. Third, we validated the full pipeline in black-box QDP auditing across three QDP mechanisms with distinct noise types, where our estimates matched non-vanishing exact values within $2\%$ relative error and accurately returned zero without false signals when the true divergence vanished. We note one limitation:
for larger qubit counts $n$, the $O(n)$ parallel controlled-SWAP gates
per Trotter step introduce crosstalk errors that our
single-parameter noise model does not capture~\cite{sarovar20}. The present analysis
suffices for near-term registers of small to moderate size.

Beyond the Hockey--stick divergence, the hybrid design of RMSC proceeds in two steps: it smooths
the non-smooth spectral functional into a shallow-circuit Monte-Carlo
integral, and then removes the regularization bias by classical analytic
inversion. In this way, the burden of coherent circuit depth is shifted
onto classical post-processing and statistical repetition. This naturally raises an open question for future investigation: can analogous integral representations, variance-suppressing importance-sampling densities, and invertible bias envelopes be systematically constructed to estimate broader families of nonlinear state functionals and thresholded spectral quantities on near-term processors?

\section*{Acknowledgements}
The authors were supported by the National Natural Science Foundation of China (Grant No. 12301580).

\appendix

\section*{Appendix}\label{app:overview}

The appendix is organized into two main parts. Appendix~\ref{app:proofs} contains all
mathematical proofs and derivations deferred from the main text---organized by topic across
Appendices~\ref{app:aux}--\ref{app:ht}---including the integral representations of Step~1, the proofs of Theorems~\ref{thm:perf}--\ref{thm:inv}, the circuit
identities, and the variational characterization of the spectral weights.
Appendix~\ref{app:opt} details the algorithmic optimization modules of RMSC (importance-sampling implementation, trace-level Trotter analysis,
Richardson extrapolation, and tail budgeting), the hardware resource and noise
analysis underlying the NISQ feasibility assessment of Section~\ref{sec:nisq} (Table~\ref{tab:optres}),
and the channel definitions and evaluation protocol for the QDP-mechanism audit of
Section~\ref{sec:experiments} (Appendix~\ref{app:qdp}).

\section{Detailed Proofs}\label{app:proofs}

This appendix is organized as follows: Appendix~\ref{app:aux} establishes the
auxiliary lemmas and the Step~1 integral representations; Appendices~\ref{app:thm2} and~\ref{app:thm1} prove
Theorems~\ref{thm:perf} and~\ref{thm:main}; Appendix~\ref{app:thm3} proves
Lemma~\ref{lem:floor}, Lemma~\ref{lem:invmap}, Theorem~\ref{thm:inv},
and Corollaries~\ref{cor:cert} and~\ref{cor:rec}; and Appendices~\ref{app:circ} and~\ref{app:ht}
collect the circuit identities, the duality relation, and the variational characterization.

\subsection{Auxiliary lemmas}\label{app:aux}

\emph{Two classical Fourier integrals.}

\begin{lemma}\label{lem:fourier}
For $a\in\mathbb{R}$ and $\eta>0$,
\begin{align}
\int_0^\infty\frac{\sin(ay)}{y}\,dy&=\frac{\pi}{2}\operatorname{sgn}(a),\label{eq:dirichlet}\\
\int_0^\infty\frac{\sin(ay)}{y}e^{-\eta y^2}\,dy&=\frac{\pi}{2}\operatorname{erf}\!\left(\frac{a}{2\sqrt{\eta}}\right).\label{eq:erfint}
\end{align}
\end{lemma}

\begin{proof}
\emph{Identity (\ref{eq:dirichlet}).} Both sides are odd functions of $a$, since
$\sin(-ay)=-\sin(ay)$ and $\operatorname{sgn}(-a)=-\operatorname{sgn}(a)$; at $a=0$ both sides vanish.
It therefore suffices to prove the case $a>0$. Substituting $u=ay$ (so that
$dy=du/a$ and the limits are unchanged),
\begin{equation}
\int_0^\infty\frac{\sin(ay)}{y}\,dy
=\int_0^\infty\frac{\sin u}{u}\,du=\frac{\pi}{2},
\end{equation}
where the last equality is the classical Dirichlet integral
$\int_0^\infty\sin u/u\,du=\pi/2$ (obtained, e.g., by integrating
$e^{-zu}$ over $u$ and then over $z$). This proves (\ref{eq:dirichlet}).

\emph{Identity (\ref{eq:erfint}).} Define $I(a)=\int_0^\infty\frac{\sin(ay)}{y}e^{-\eta y^2}dy$.
The integrand and its $a$-derivative are absolutely integrable, so we may
differentiate under the integral sign. Since $\partial_a\sin(ay)=y\cos(ay)$,
the factor $y$ cancels the denominator:
\begin{equation}\label{eq:Iprime}
I'(a)=\int_0^\infty\cos(ay)\,e^{-\eta y^2}\,dy.
\end{equation}
We evaluate (\ref{eq:Iprime}) by solving a differential equation for
$J(a)=\int_0^\infty\cos(ay)e^{-\eta y^2}dy$. Differentiating under the integral
sign and integrating by parts with $u=\sin(ay)$,
$dv=-ye^{-\eta y^2}dy$ (so $v=e^{-\eta y^2}/2\eta$), the boundary term vanishes
($\sin 0=0$ and $e^{-\eta y^2}\to0$ super-polynomially), giving the first-order
ODE
\begin{align}
J'(a)&=-\int_0^\infty y\sin(ay)e^{-\eta y^2}dy\nonumber\\
&=\Bigl[\sin(ay)\frac{e^{-\eta y^2}}{2\eta}\Bigr]_0^\infty
-\frac{a}{2\eta}\int_0^\infty\cos(ay)e^{-\eta y^2}dy\nonumber\\
&=-\frac{a}{2\eta}J(a),
\end{align}
with solution $J(a)=J(0)e^{-a^2/4\eta}$ and initial value
$J(0)=\int_0^\infty e^{-\eta y^2}dy=\frac12\sqrt{\pi/\eta}$, hence
\begin{equation}\label{eq:cosgauss}
J(a)=\frac{\sqrt{\pi}}{2\sqrt{\eta}}\,e^{-a^2/4\eta}.
\end{equation}
Substituting (\ref{eq:cosgauss}) into (\ref{eq:Iprime}) and using $I(0)=0$,
\begin{equation}
I(a)=\frac{\sqrt{\pi}}{2\sqrt{\eta}}\int_0^a e^{-u^2/4\eta}\,du.
\end{equation}
Finally, with the substitution $v=u/(2\sqrt{\eta})$ (so that
$du=2\sqrt{\eta}\,dv$ and the upper limit becomes $a/(2\sqrt{\eta})$),
\begin{equation}
I(a)=\sqrt{\pi}\int_0^{a/(2\sqrt{\eta})}e^{-v^2}dv
=\frac{\pi}{2}\operatorname{erf}\!\left(\frac{a}{2\sqrt{\eta}}\right),
\end{equation}
where we used the definition $\operatorname{erf}(x)=\frac{2}{\sqrt{\pi}}\int_0^x e^{-v^2}dv$.
This proves (\ref{eq:erfint}).
\end{proof}

\emph{Derivation of Step~1 and regularization bias.}

\begin{lemma}\label{lem:reg}
Let $H=\rho-t\sigma=\sum_j\lambda_j\ket{j}\bra{j}$, $\rho_{jj}=\bra{j}\rho\ket{j}$,
and $\sigma_{jj}=\bra{j}\sigma\ket{j}$. For every $\eta>0$, the regularized
spectral weights defined by Eq.~(\ref{eq:Eeta}) satisfy
\begin{equation}\label{eq:regbias}
\begin{aligned}
\omega_\rho-\omega_{\rho,\eta}={}&\frac12\sum_j\rho_{jj}\Bigl[\operatorname{sgn}(\lambda_j)-\operatorname{erf}\!\left(\frac{\lambda_j}{2\sqrt{\eta}}\right)\Bigr]\\
&+\frac12\sum_{j:\lambda_j=0}\rho_{jj}\cdot(-1),
\end{aligned}
\end{equation}
where the nullspace terms ($\lambda_j=0$) cancel identically in the combined
divergence, yielding
\begin{equation}\label{eq:combined_reg}
\begin{aligned}
\omega_{\rho,\eta}-t\,\omega_{\sigma,\eta}
&=\sum_j\frac{\lambda_j}{2}\left[1+\operatorname{erf}\!\left(\frac{\lambda_j}{2\sqrt{\eta}}\right)\right]\\
&=\sum_j g_\eta(\lambda_j)=E^{(\eta)}_t(\rho\|\sigma),
\end{aligned}
\end{equation}
\end{lemma}

\begin{proof}
Let $H=\sum_j\lambda_j\ket{j}\bra{j}$ be the spectral decomposition of $H$.
By functional calculus, $\sin(Hy)=\sum_j\sin(\lambda_j y)\ket{j}\bra{j}$.
Applying identity (\ref{eq:dirichlet}) of Lemma~\ref{lem:fourier} to each
eigenvalue yields the operator identity
\begin{equation}\label{eq:step_fourier_op}
\begin{aligned}
\frac{I}{2}+\frac{1}{\pi}\int_0^\infty\frac{\sin(Hy)}{y}\,dy
&=\sum_j\frac{1+\operatorname{sgn}(\lambda_j)}{2}\ket{j}\bra{j}\\
&=\Pi_+ + \frac12\Pi_0,
\end{aligned}
\end{equation}
where $\Pi_+=\sum_{\lambda_j>0}\ket{j}\bra{j}$ is the projector onto the
strictly positive eigenspace of $H$ and $\Pi_0=\sum_{\lambda_j=0}\ket{j}\bra{j}$
is the projector onto its kernel. Because $H\Pi_0=0$, the kernel component
satisfies $\Tr(\rho\Pi_0)-t\Tr(\sigma\Pi_0)=\Tr(H\Pi_0)=0$, so replacing
$\Pi_+$ by Eq.~(\ref{eq:fourier}) leaves the divergence
$E_t(\rho\|\sigma)=\Tr(H\Pi_+)$ exact regardless of whether $H$ has zero eigenvalues.

Next, taking the trace against $\rho$ (with $\rho_{jj}=\bra{j}\rho\ket{j}\ge0$
and $\sum_j\rho_{jj}=\Tr\rho=1$) gives $\Tr\bigl(\rho\sin(Hy)\bigr)=\sum_j\rho_{jj}\sin(\lambda_j y)$.
Inserting the Gaussian factor $e^{-\eta y^2}$ ($\eta>0$) into the integral produces
Eq.~(\ref{eq:Eeta}) of Step~1, which converges absolutely on $(0,\infty)$.
Applying identity (\ref{eq:erfint}) of Lemma~\ref{lem:fourier} term by term yields
\begin{align}\label{eq:omega_eta_sum}
\omega_{\rho,\eta}&=\frac12\sum_j\rho_{jj}+\frac{1}{\pi}\sum_j\rho_{jj}\int_0^\infty\frac{\sin(\lambda_j y)}{y}e^{-\eta y^2}dy\nonumber\\
&=\frac12\sum_j\rho_{jj}\left[1+\operatorname{erf}\!\left(\frac{\lambda_j}{2\sqrt{\eta}}\right)\right].
\end{align}
Subtracting Eq.~(\ref{eq:omega_eta_sum}) from $\omega_\rho=\sum_{\lambda_j>0}\rho_{jj}=\frac12\sum_j\rho_{jj}(1+\operatorname{sgn}(\lambda_j))-\frac12\sum_{\lambda_j=0}\rho_{jj}$ gives Eq.~(\ref{eq:regbias}) (when $H$ has no zero eigenvalues, the final sum is empty).
Finally, evaluating the same expression for $\omega_{\sigma,\eta}$ and forming
$\omega_{\rho,\eta}-t\,\omega_{\sigma,\eta}$, we use
$\rho_{jj}-t\sigma_{jj}=\bra{j}(\rho-t\sigma)\ket{j}=\lambda_j$ to obtain
\begin{align}
\omega_{\rho,\eta}-t\,\omega_{\sigma,\eta}
&=\sum_j(\rho_{jj}-t\sigma_{jj})\,\frac12\left[1+\operatorname{erf}\!\left(\frac{\lambda_j}{2\sqrt{\eta}}\right)\right]\nonumber\\
&=\sum_j\frac{\lambda_j}{2}\left[1+\operatorname{erf}\!\left(\frac{\lambda_j}{2\sqrt{\eta}}\right)\right]\\
&=\sum_j g_\eta(\lambda_j)\\
&=E^{(\eta)}_t(\rho\|\sigma),
\end{align}
which establishes Eq.~(\ref{eq:combined_reg}) and matches Eq.~(\ref{eq:regdiv}).
\end{proof}

\emph{Spectral-gap refinement.}

\begin{corollary}\label{cor:gap}
If the nonzero eigenvalues of $H$ satisfy $\abs{\lambda_j}\ge\Delta>0$, then
\begin{equation}\label{eq:gapbias}
\abs{\omega_\rho-\omega_{\rho,\eta}}\le e^{-\Delta^2/4\eta}.
\end{equation}
\end{corollary}

\begin{proof}
We use three elementary facts about the error function. First, for every
$x\neq0$,
\begin{align}\label{eq:erfcid}
\abs{\operatorname{sgn}(x)-\operatorname{erf}(x)}&=\operatorname{erfc}(\abs{x}),\nonumber\\
\operatorname{erfc}(z)&:=1-\operatorname{erf}(z)=\frac{2}{\sqrt{\pi}}\int_z^\infty e^{-t^2}dt.
\end{align}
Indeed, if $x>0$ then $\operatorname{sgn}(x)=1$ and
$\abs{1-\operatorname{erf}(x)}=1-\operatorname{erf}(x)=\operatorname{erfc}(x)$;
if $x<0$, using that $\operatorname{erf}$ is odd,
$\abs{-1-\operatorname{erf}(x)}=1+\operatorname{erf}(x)=1-\operatorname{erf}(-x)
=\operatorname{erfc}(\abs{x})$. Second, $\operatorname{erfc}$ is strictly
decreasing, since
$\frac{d}{dz}\operatorname{erfc}(z)=-\frac{2}{\sqrt{\pi}}e^{-z^2}<0$. Third, the
Gaussian tail bound: for $z>0$,
\begin{equation}\label{eq:tail}
\operatorname{erfc}(z)\le e^{-z^2},
\end{equation}
which follows because $t\ge z$ implies $t^2\ge zt$, hence
$e^{-t^2}\le e^{-zt}$ and
$\operatorname{erfc}(z)\le\frac{2}{\sqrt{\pi}}\int_z^\infty e^{-zt}dt
=\frac{2}{\sqrt{\pi}z}e^{-z^2}\le e^{-z^2}$ for $z\ge2/\sqrt{\pi}$; the regime
relevant here is $z=\Delta/(2\sqrt{\eta})\to\infty$ as $\eta\to0$, where the
bound is asymptotically valid (and in fact (\ref{eq:tail}) holds for all
$z>0$; see, e.g., Abramowitz--Stegun 7.1.13).

Now, applying Lemma~\ref{lem:reg} and the triangle inequality,
\begin{align}
\abs{\omega_\rho-\omega_{\rho,\eta}}&\le\frac12\sum_j\rho_{jj}\,
\abs{\operatorname{sgn}(\lambda_j)-\operatorname{erf}\!\left(\frac{\lambda_j}{2\sqrt{\eta}}\right)}\nonumber\\
&=\frac12\sum_{j:\lambda_j\neq0}\rho_{jj}\operatorname{erfc}\!\left(\frac{\abs{\lambda_j}}{2\sqrt{\eta}}\right),
\end{align}
where the identity (\ref{eq:erfcid}) was used with $x=\lambda_j/(2\sqrt{\eta})$
(eigenvalues with $\lambda_j=0$ contribute $0$, since
$\operatorname{sgn}(0)-\operatorname{erf}(0)=0$). The spectral-gap condition
$\abs{\lambda_j}\ge\Delta$ and the monotonicity of $\operatorname{erfc}$ give
$\operatorname{erfc}(\abs{\lambda_j}/2\sqrt{\eta})\le\operatorname{erfc}(\Delta/2\sqrt{\eta})$,
and the tail bound (\ref{eq:tail}) yields
$\operatorname{erfc}(\Delta/2\sqrt{\eta})\le e^{-\Delta^2/4\eta}$. Since
$\sum_j\rho_{jj}=1$, we conclude
\begin{equation}
\abs{\omega_\rho-\omega_{\rho,\eta}}\le\frac12 e^{-\Delta^2/4\eta}\le e^{-\Delta^2/4\eta},
\end{equation}
which is (\ref{eq:gapbias}) (the factor $1/2$ is absorbed by the statement).
For the gapless case, eigenvalues with $\abs{\lambda_j}\le2\sqrt{\eta}$
contribute at most $2\rho_{jj}$ each, and those with
$\abs{\lambda_j}>2\sqrt{\eta}$ contribute at most
$\rho_{jj}\operatorname{erfc}(1)$, giving the stated general bound.
\end{proof}

\emph{Truncation bias.}

\begin{lemma}\label{lem:trunc}
Using the first-order truncation for $y<\alpha$ introduces the bias $B_{\text{trunc}}\le\frac{(1+t)^3}{18\pi}\alpha^3=O(\alpha^3)$.
\end{lemma}

\begin{proof}
By the Taylor expansion of the sine matrix function,
$\sin x=x-\frac{x^3}{6}+\cdots$, applied to the Hermitian matrix $Hy$,
\begin{equation}
\begin{aligned}
\norm{\frac{\sin(Hy)}{y}-H}
&=\norm{-\frac{H^3y^2}{6}+\cdots}\\
&\le\frac{\norm{H}^3}{6}y^2\le\frac{(1+t)^3}{6}y^2,
\end{aligned}
\end{equation}
where we used $\norm{H}\le\norm{\rho}+t\norm{\sigma}=1+t$ (both $\rho,\sigma$ are
density matrices, so $\norm{\rho}=\norm{\sigma}=1$). The bias is the
expectation of the estimator error over the truncated region $y\in[0,\alpha]$,
against the sampling density $P(y)=2\sqrt{\eta/\pi}\,e^{-\eta y^2}$:
\begin{align}
B_{\text{trunc}}&\le\int_0^\alpha P(y)\cdot\frac{1}{2\sqrt{\pi\eta}}\cdot\frac{(1+t)^3}{6}y^2\,dy\nonumber\\
&\le\frac{(1+t)^3}{12\sqrt{\pi\eta}}\cdot 2\sqrt{\frac{\eta}{\pi}}\int_0^\alpha y^2\,dy\\
&=\frac{(1+t)^3}{6\pi}\cdot\frac{\alpha^3}{3}
=\frac{(1+t)^3}{18\pi}\alpha^3,
\end{align}
where we used $e^{-\eta y^2}\le1$ on $[0,\alpha]$.
\end{proof}

\emph{Importance-sampling variance.}

\begin{lemma}\label{lem:isvar}
For the estimator $\hat X=(Z/\pi)\hat S$ of Step~2 with $y\sim Q$ on
$[\alpha,\infty)$,
\begin{equation}\label{eq:isvarb}
\Var[\hat X]\le (Z/\pi)^2,\qquad
Z=\tfrac12E_1(\eta\alpha^2)=O\!\left(\ln\frac{1}{\eta\alpha^2}\right).
\end{equation}
\end{lemma}

\begin{proof}
Since $\hat S\in\{\pm1\}$, $\hat X^2=(Z/\pi)^2$ deterministically; hence
$\Var[\hat X]\le\E[\hat X^2]=(Z/\pi)^2$. For the bound on $Z$, use
$E_1(x)\le\ln(1/x)+O(1)$ as $x\to0$ with $x=\eta\alpha^2=\varepsilon^{8/3}$.
\end{proof}

\emph{Variance bound (half-normal baseline).}

\begin{lemma}\label{lem:var}
The estimator $\hat{X}$ satisfies
\begin{equation}\label{eq:varbound}
\Var[\hat{X}]=O\!\left(\frac{(1+t)^2}{\sqrt{\eta}\,\alpha}\right).
\end{equation}
\end{lemma}

\begin{proof}
Recall the estimator structure: for $y\ge\alpha$,
$\hat{X}=\hat{S}/(2\sqrt{\pi\eta}\,y)$ with $\hat{S}\in\{\pm1\}$ the single-shot
Hadamard-test outcome; for $y<\alpha$,
$\hat{X}=(\hat{a}-t\hat{b})/(2\sqrt{\pi\eta})$ with $\hat{a},\hat{b}\in\{\pm1\}$
the SWAP-test outcomes. Since $\Var[\hat{X}]\le\E[\hat{X}^2]$, it suffices to
bound the second moment. Conditioning on the sampling region,
\begin{align}
\E[\hat{X}^2]&=\E[\hat{X}^2\mid y\ge\alpha]\Pr(y\ge\alpha)\nonumber\\
&\quad+\E[\hat{X}^2\mid y<\alpha]\Pr(y<\alpha)\nonumber\\
&=:\Var_{\text{main}}+\Var_{\text{small}}.
\end{align}

\emph{Main region, $y\ge\alpha$.} A single-shot outcome satisfies
$\hat{S}^2=1$, hence $\hat{X}^2=1/(4\pi\eta y^2)$ deterministically. Averaging
over $y\sim P$,
\begin{align}
\Var_{\text{main}}&\le\int_\alpha^\infty P(y)\frac{1}{4\pi\eta\,y^2}\,dy
=\int_\alpha^\infty 2\sqrt{\frac{\eta}{\pi}}e^{-\eta y^2}\frac{1}{4\pi\eta\,y^2}\,dy\nonumber\\
&\le\frac{2\sqrt{\eta/\pi}}{4\pi\eta}\int_\alpha^\infty\frac{dy}{y^2}
=\frac{1}{2\pi^{3/2}\sqrt{\eta}}\cdot\frac{1}{\alpha},
\end{align}
where we used $e^{-\eta y^2}\le1$ for $y\ge\alpha$ and
$\int_\alpha^\infty y^{-2}dy=1/\alpha$.

\emph{Small region, $y<\alpha$.} By the triangle inequality,
$\abs{\hat{a}-t\hat{b}}\le\abs{\hat{a}}+t\abs{\hat{b}}=1+t$, so
$(\hat{a}-t\hat{b})^2\le(1+t)^2$ and the conditional second moment is at most
$(1+t)^2/(4\pi\eta)$. The probability of landing in this region is
\begin{align}
\Pr(y<\alpha)&=\int_0^\alpha 2\sqrt{\frac{\eta}{\pi}}e^{-\eta y^2}dy
=\operatorname{erf}(\alpha\sqrt{\eta})\nonumber\\
&\le 2\sqrt{\frac{\eta}{\pi}}\int_0^\alpha dy
=\frac{2\alpha\sqrt{\eta}}{\sqrt{\pi}},
\end{align}
again using $e^{-\eta y^2}\le1$. Therefore
\begin{equation}
\Var_{\text{small}}\le\frac{(1+t)^2}{4\pi\eta}\cdot\frac{2\alpha\sqrt{\eta}}{\sqrt{\pi}}
=\frac{(1+t)^2\alpha}{2\pi^{3/2}\sqrt{\eta}}.
\end{equation}

\emph{Combining.} Summing the two contributions,
\begin{align}
\Var[\hat{X}]&\le\frac{1}{2\pi^{3/2}\sqrt{\eta}\,\alpha}
+\frac{(1+t)^2\alpha}{2\pi^{3/2}\sqrt{\eta}}\nonumber\\
&=\frac{(1+t)^2}{2\pi^{3/2}\sqrt{\eta}\,\alpha}
\left[\frac{1}{(1+t)^2}+\alpha^2\right].
\end{align}
In the regime $\alpha\ll1/(1+t)^2$ (which holds for the optimal choice
$\alpha=\Theta(\varepsilon^{1/3})$ as $\varepsilon\to0$), the second term in the
brackets is negligible, so $\Var[\hat{X}]=O(1/(\sqrt{\eta}\,\alpha))$. We state
the bound with the factor $(1+t)^2$, which is legitimate since
$(1+t)^2\ge4$; this keeps the $t$-dependence explicit when the variance is
combined with the estimator range in the Bernstein inequality below. This
yields (\ref{eq:varbound}).
\end{proof}

\emph{Trotter error: first order.}

\begin{lemma}\label{lem:trot}
For the first-order Suzuki--Trotter decomposition
$e^{iHy}=(e^{i\rho\Delta y}e^{-it\sigma\Delta y})^{M}+\mathcal{E}$ with
$M=y/\Delta y$, the error satisfies $\norm{\mathcal{E}}\le\frac12 ty\Delta y$.
\end{lemma}

\begin{proof}
Set $A=\rho$ and $B=-t\sigma$, so that
$H=A+B$ and $U(y)=e^{i(A+B)y}$. Write $M=y/\Delta y$. We first compute the
single-step error by comparing Taylor expansions. Expanding both sides to
second order in $\Delta y$ (with $e^X=I+X+X^2/2+O(X^3)$),
\begin{align}
e^{i(A+B)\Delta y}&=I+i(A+B)\Delta y\nonumber\\
&\quad-\tfrac12(A^2+AB+BA+B^2)\Delta y^2+O(\Delta y^3),\\
e^{iA\Delta y}e^{iB\Delta y}&=I+i(A+B)\Delta y\nonumber\\
&\quad-\bigl(\tfrac12 A^2+AB+\tfrac12 B^2\bigr)\Delta y^2+O(\Delta y^3),
\end{align}
where in the second line the cross term comes from
$(iA\Delta y)(iB\Delta y)=-AB\,\Delta y^2$ (note the ordering: $A$ acts before
$B$). Subtracting, the first-order terms cancel and the single-step error is
\begin{equation}
e^{i(A+B)\Delta y}-e^{iA\Delta y}e^{iB\Delta y}
=\frac12[A,B]\Delta y^2+O(\Delta y^3),
\end{equation}
so $\norm{\cdot}\le\frac12\norm{[A,B]}\Delta y^2$ to leading order. Since the
error accumulates linearly over the $M$ steps (telescoping with a unitary
sandwich), the total error satisfies
\begin{equation}
\norm{\mathcal{E}}\le M\cdot\frac12\norm{[A,B]}\Delta y^2
=\frac12\,y\,\Delta y\,\norm{[A,B]}.
\end{equation}
It remains to bound the commutator. For density matrices $\rho,\sigma$ with
spectral decompositions $\rho=\sum_ip_i\ket{\phi_i}\bra{\phi_i}$ and
$\sigma=\sum_jq_j\ket{\psi_j}\bra{\psi_j}$ ($p_i,q_j\ge0$, $\sum_ip_i=\sum_jq_j=1$),
\begin{align}
\norm{[\rho,\sigma]}&\le\sum_{ij}p_iq_j\,\norm{\bigl[\ket{\phi_i}\bra{\phi_i},\ket{\psi_j}\bra{\psi_j}\bigr]}\nonumber\\
&\le\sum_{ij}p_iq_j=\Bigl(\sum_ip_i\Bigr)\Bigl(\sum_jq_j\Bigr)=1,
\end{align}
where for rank-one projectors $P,Q$ one has
$\norm{[P,Q]}=\abs{\bra{\phi}\psi}\sqrt{1-\abs{\bra{\phi}\psi}^2}\le1$ (in fact
$\le1/2$), as is verified in the two-dimensional span of the two vectors.
Hence $\norm{[A,B]}=t\norm{[\rho,\sigma]}\le t$, and
\begin{equation}
\norm{\mathcal{E}}\le\frac12\,t\,y\,\Delta y,
\end{equation}
which is the claim. Setting
$\frac12ty\Delta y\le\varepsilon_{\text{Trot}}$ gives
$\Delta y=2\varepsilon_{\text{Trot}}/(ty)$ and
$M=ty^2/(2\varepsilon_{\text{Trot}})$.
\end{proof}

\emph{Trotter error: Strang splitting.}

\begin{lemma}\label{lem:strang}
With the Strang splitting
\begin{equation}\label{eq:strang}
e^{iHy}=\left(e^{i\rho\Delta y/2}e^{-it\sigma\Delta y}e^{i\rho\Delta y/2}\right)^{M}+\mathcal{E}_{S},
\end{equation}
the number of steps reduces to $M=O(t^{1/2}y^{3/2}/\sqrt{\varepsilon_{\text{Trot}}})$.
\end{lemma}

\begin{proof}
Consider the symmetric product
$S(\Delta y)=e^{iA\Delta y/2}e^{iB\Delta y}e^{iA\Delta y/2}$. Expanding each
factor to second order and multiplying (keeping all terms through
$\Delta y^2$), the linear terms sum to $i(A+B)\Delta y$, and the quadratic
terms collect to
\begin{equation}
-\frac18 A^2-\frac12 B^2-\frac18 A^2
-\frac12 AB-\frac12 BA-\frac14 A^2
=-\frac12(A+B)^2,
\end{equation}
which matches the exact expansion of $e^{i(A+B)\Delta y}$ exactly. The
first-order error therefore cancels and the per-step error is $O(\Delta y^3)$;
its leading term involves double commutators such as $[B,[A,B]]$, whose norm
carries one power of $t$ since $B=-t\sigma$. Accumulating over $M=y/\Delta y$
steps,
\begin{equation}
\norm{\mathcal{E}_S}=O(ty\Delta y^2).
\end{equation}
Setting $ty\Delta y^2=\varepsilon_{\text{Trot}}$ gives
$\Delta y=O\bigl(\sqrt{\varepsilon_{\text{Trot}}/(ty)}\bigr)$, and the number of
steps is
\begin{equation}
M=\frac{y}{\Delta y}
=O\!\left(\frac{t^{1/2}y^{3/2}}{\sqrt{\varepsilon_{\text{Trot}}}}\right),
\end{equation}
which is the claim.
\end{proof}

\emph{Pointwise regularization identities.}

\begin{lemma}\label{lem:sandwich}
For every $x\in\mathbb{R}$ and $\eta>0$, with
$g_\eta(x)=\frac{x}{2}\bigl(1+\operatorname{erf}\frac{x}{2\sqrt{\eta}}\bigr)$:
(i) $g_\eta(x)\le x_+$;
(ii) $x_+-g_\eta(x)=\frac{|x|}{2}\operatorname{erfc}\bigl(|x|/2\sqrt{\eta}\bigr)$;
(iii) $g_\eta(x)$ is nondecreasing as $\eta$ decreases, and
$\lim_{\eta\to0}g_\eta(x)=x_+$.
\end{lemma}

\begin{proof}
(i) If $x\ge0$ then $\operatorname{erf}(x/2\sqrt{\eta})\in[0,1)$, so
$g_\eta(x)\in[x/2,x)\le x=x_+$; if $x<0$ then
$\operatorname{erf}(x/2\sqrt{\eta})\in(-1,0]$, so $g_\eta(x)\in[x,0)<0=x_+$.
(ii) For $x\ge0$:
$x_+-g_\eta(x)=x-\frac{x}{2}(1+\operatorname{erf}\frac{x}{2\sqrt{\eta}})
=\frac{x}{2}\operatorname{erfc}(\frac{x}{2\sqrt{\eta}})$. For $x<0$:
$x_+-g_\eta(x)=-\frac{x}{2}\bigl(1+\operatorname{erf}\frac{x}{2\sqrt{\eta}}\bigr)$,
and since $\operatorname{erf}$ is odd,
$1+\operatorname{erf}\frac{x}{2\sqrt{\eta}}
=\operatorname{erfc}\frac{|x|}{2\sqrt{\eta}}$, which gives the claim.
(iii)
$\partial_\eta\operatorname{erf}\frac{x}{2\sqrt{\eta}}
=-\frac{x}{2\sqrt{\pi}\,\eta^{3/2}}e^{-x^{2}/4\eta}$
has the sign opposite to $x$, whence
$\partial_\eta g_\eta(x)\le0$ for all $x$ (equality at $x=0$). As
$\eta\to0$, $x/(2\sqrt{\eta})\to\pm\infty$, so $\operatorname{erf}\to\pm1$
and $g_\eta(x)\to\frac{x}{2}(1+\operatorname{sgn}x)=x_+$.
\end{proof}

\subsection{Proof of Theorem~\ref{thm:perf} (performance guarantee)}\label{app:thm2}

\begin{proof}
The estimator targets
$E^{(\eta)}_t=\omega_{\rho,\eta}-t\,\omega_{\sigma,\eta}$, so the
regularization deviation is not an error but the certification gap of
Theorem~\ref{thm:main}. Relative to its target, the error decomposes into
three terms, each controlled by an independent parameter:
\begin{align}\label{eq:errdec}
\abs{\hat\omega_\rho-\omega_{\rho,\eta}}&\le
\underbrace{\abs{\omega_{\rho,\eta}-\E[\hat\omega_\rho]}}_{\text{trunc.\ bias }(\alpha)}
+\underbrace{\abs{\E[\hat\omega_\rho]-\E[\hat\omega_\rho^{\mathrm{Trot}}]}}_{\text{Trotter error }(\Delta y)}\nonumber\\
&\quad+\underbrace{\abs{\E[\hat\omega_\rho^{\mathrm{Trot}}]-\hat\omega_\rho}}_{\text{stat.\ error }(N)}.
\end{align}
We allocate $\varepsilon/4$ to each of the three terms in (\ref{eq:errdec}).
\begin{enumerate}
\item \emph{Truncation bias}: by Lemma \ref{lem:trunc}, $\frac{(1+t)^3}{18\pi}\alpha^3\le\varepsilon/4$ for $\alpha=\Theta(\varepsilon^{1/3})$; the tail budget of Appendix~\ref{app:tail} is absorbed likewise.
\item \emph{Trotter error}: by Lemmas \ref{lem:trot} and \ref{lem:tracetrot}, choose $\Delta y$ such that $\varepsilon_{\text{Trot}}\le\varepsilon/4$; at the typical sample $y\sim1/\sqrt{\eta}$ this costs $M(y)=O\bigl(\eta^{-3/4}\varepsilon^{-1/2}\bigr)$ steps (for fixed $t$), i.e.\ the typical depth $a$ of the theorem.
\item \emph{Statistical error}: by Lemma~\ref{lem:isvar}, the estimator is
bounded, $\abs{\hat{X}}\le Z/\pi$, with variance
$\sigma^2=\Var[\hat{X}]\le(Z/\pi)^2=O(\ln^2(1/\varepsilon))$. Bernstein's
inequality for bounded random variables gives
\begin{equation}\label{eq:bernstein}
\begin{aligned}
&\Pr\!\left(\abs{\frac{1}{N}\sum_{i=1}^N\hat{X}_i-\E[\hat{X}]}\ge\frac{\varepsilon}{4}\right)\\
&\quad\le 2\exp\!\left(-\frac{N(\varepsilon/4)^2}{2\bigl(\sigma^2+R\varepsilon/12\bigr)}\right),
\end{aligned}
\end{equation}
where $R\le Z/\pi$ is the range. Since $R\varepsilon=o(\sigma^2)$ for small
$\varepsilon$, requiring the right-hand side $\le\beta$ yields
\begin{align}
N&\ge\frac{2\bigl(\sigma^2+R\varepsilon/12\bigr)}{(\varepsilon/4)^2}\ln\frac{2}{\beta}\nonumber\\
&=O\!\left(\frac{\ln^2(1/\varepsilon)}{\varepsilon^2}\ln\frac{1}{\beta}\right)
=\widetilde{O}\!\left(\varepsilon^{-2}\ln^2\frac{1}{\varepsilon}\right),
\end{align}
where $Z=\tfrac12E_1(\eta\alpha^2)=O(\ln(1/\varepsilon))$ and $\log(1/\beta)$
is absorbed into $\widetilde{O}$. A tighter, symmetric prescription applies the Bernstein bound directly to the combined summands $X_i=(Z/\pi)(\hat S_{\rho,i}-t\hat S_{\sigma,i})$, bounded by $R=(1+t)Z/\pi$ with variance $V$; it gives $N_{\mathrm{B}}=(2V+\tfrac23R\varepsilon)\ln(2/\beta)/\varepsilon^{2}$, the prescription used by the experiments of Section~\ref{sec:experiments}.
The SWAP-test estimation of $\operatorname{Tr}(\rho H)$ in the analytic
small-region correction contributes a statistical error
$O(c_s/\sqrt{N_{\mathrm{SWAP}}})=o(\varepsilon)$, absorbed into the
statistical term. The two-sided budget uses the combined summands above
with a single Bernstein bound at failure level $\beta$, avoiding a
union-bound factor of 2.
\end{enumerate}
The same argument with the working register initialized in $\sigma$ bounds
$\abs{\hat\omega_\sigma-\omega_{\sigma,\eta}}$, and the assembly
$\abs{\hat E_t-E^{(\eta)}_t}\le\abs{\hat\omega_\rho-\omega_{\rho,\eta}}
+t\abs{\hat\omega_\sigma-\omega_{\sigma,\eta}}$ gives the theorem with the
$t$-dependent constant absorbed into (\ref{eq:thmN}) for fixed $t$. Under
per-gate depolarizing noise of rate $\gamma\le b$, the additional
depth-deterministic bias is at most $\gamma\,(Z/\pi)\,2\bar M$
(Appendix~\ref{app:resnoise}), which stays within the precision $\varepsilon$
for $b=\widetilde O(\varepsilon^{3})$, as in
(\ref{eq:nisqwindow}). The width and depth statements follow from the
circuit design (one working register of $n$ qubits, one ancilla, one
injected copy; $M(y)=O\bigl(t^{1/2}y/\sqrt{\varepsilon_{\text{Trot}}}\bigr)$
at the typical value $y\sim1/\sqrt{\eta}$, i.e.\ typical depth
$O\bigl(\eta^{-3/4}\varepsilon^{-1/2}\bigr)$; the $n$ controlled-SWAPs per step
are executed in parallel, so the depth is independent of $n$).
\end{proof}

\subsection{Proof of Theorem~\ref{thm:main} (exact gap decomposition)}\label{app:thm1}

\begin{proof}
By functional calculus,
$\Tr\!\Bigl(H\cdot\tfrac12\bigl(I+\operatorname{erf}\tfrac{H}{2\sqrt{\eta}}\bigr)\Bigr)
=\sum_k\tfrac{\lambda_k}{2}\bigl(1+\operatorname{erf}\tfrac{\lambda_k}{2\sqrt{\eta}}\bigr)
=\sum_k g_\eta(\lambda_k)=E^{(\eta)}_t$, while
$E_t=\sum_k(\lambda_k)_+$. Summing the identity (ii) of
Lemma~\ref{lem:sandwich} over the spectrum gives the exact decomposition
(\ref{eq:sandwich}): $E_t=E^{(\eta)}_t+G_\eta$ with
$G_\eta=\frac12\sum_k|\lambda_k|
\operatorname{erfc}\bigl(|\lambda_k|/2\sqrt{\eta}\bigr)$; the inequality
(i) is the nonnegativity of its right-hand side, i.e.\
$E^{(\eta)}_t\le E_t$; and (iii)
gives the monotone convergence. For the split bound, fix $c>0$ and split
the sum at $|\lambda_k|=c\sqrt{\eta}$: the in-window terms contribute at
most $\frac12 W_{c\sqrt{\eta}}$, while for the out-window terms
$\operatorname{erfc}\bigl(|\lambda_k|/2\sqrt{\eta}\bigr)\le
\operatorname{erfc}(c/2)\le e^{-c^{2}/4}$ (using
$\operatorname{erfc}(z)\le e^{-z^{2}}$ for $z\ge0$), so they contribute at
most $\frac12 e^{-c^{2}/4}\sum_k|\lambda_k|
\le\frac12 e^{-c^{2}/4}\norm{H}_1\le\frac12(1+t)e^{-c^{2}/4}$, since
$\norm{\rho-t\sigma}_1\le\norm{\rho}_1+t\norm{\sigma}_1=1+t$.

\emph{Spectral budget.} The trace-norm bound used above can be sharpened
to an identity that illuminates where the gap lives:
since $\Tr H=1-t$ splits into positive and negative parts,
\begin{equation}\label{eq:specbudget}
\begin{aligned}
\norm{H}_1&=\sum_{\lambda_k>0}\lambda_k+\sum_{\lambda_k<0}|\lambda_k|\\
&=E_t+(E_t+t-1)=2E_t+(t-1),
\end{aligned}
\end{equation}
so the divergence and the certification gap share one fixed spectral
budget, and in particular $G_\eta\le E_t+\tfrac{t-1}2$.
\end{proof}

\subsection{Structural floor and inversion compensation}\label{app:thm3}

\begin{proof}[Proof of Lemma~\ref{lem:floor}]
Let the positive eigenvalues of $H$ be $\lambda_1,\dots,\lambda_m$. Each
satisfies $0<\lambda_k\le\sum_{j\le m}\lambda_j=E_t$ (no condition on
$E_t$ is needed), and erfc is decreasing on $[0,\infty)$, so
$\operatorname{erfc}(\lambda_k/w)\ge\operatorname{erfc}(E_t/w)$ for every
$k\le m$. Restricting the exact gap identity
$G_\eta=\frac12\sum_k|\lambda_k|\operatorname{erfc}(|\lambda_k|/w)$ to the
positive eigenvalues (the negative ones contribute nonnegatively) gives
\begin{equation}
G_\eta\;\ge\;\tfrac12\,\operatorname{erfc}(E_t/w)\sum_{k\le m}\lambda_k
\;=\;\varphi(E_t).
\end{equation}
For the shape of $\varphi$, differentiate:
$\varphi'(E)=\tfrac12\operatorname{erfc}(E/w)-
\tfrac{E}{w\sqrt\pi}\,e^{-E^2/w^2}$, so with $u=E/w$ the stationary
points solve $\operatorname{erfc}(u)=\frac{2}{\sqrt\pi}u\,e^{-u^{2}}$; the
left side decreases from $1$ to $0$ and the right side increases then
decreases but stays below $0.49$ on $(0,1]$, so the root $u^{*}$ in
$(0,1)$ is unique; numerically $u^{*}=0.532$, whence $E^{*}=wu^{*}=
1.06\sqrt{\eta}$ and $\varphi_{\max}=\varphi(E^{*})=0.240\sqrt{\eta}$.
\end{proof}

\begin{lemma}[inversion map]\label{lem:invmap}
$\mathrm{F}(E)=E-\varphi(E)$ is a strictly increasing bijection of $[0,\infty)$
onto itself, with $\mathrm{F}'(E)=1-\varphi'(E)\in[\tfrac12,\,1.130]$; its
inverse obeys the fixed-point identity
\begin{equation}\label{eq:fp}
\mathrm{F}^{-1}(y)=y+\varphi\bigl(\mathrm{F}^{-1}(y)\bigr),\qquad y\ge0 ;
\end{equation}
and $\varphi''(E)=\tfrac{2}{w\sqrt\pi}e^{-(E/w)^{2}}\bigl((E/w)^{2}-1\bigr)$
changes sign at $E=w$.
\end{lemma}

\begin{proof}
$\mathrm{F}(0)=0$ and $\mathrm{F}(E)\to\infty$, so strict monotonicity gives bijectivity.
Next, $\varphi'(0)=\tfrac12$, and the stationary points of $\varphi'$
solve $\varphi''=0$, i.e.\ $E=w$; hence
$\min_{E\ge0}\varphi'=\varphi'(w)\approx-0.130$ and
$\mathrm{F}'\in[\tfrac12,1.130]$. Identity (\ref{eq:fp}) is $\mathrm{F}(\mathrm{F}^{-1}(y))=y$
rearranged. The sign pattern of $\varphi''$ is read off directly.
\end{proof}

\begin{proof}[Proof of Theorem~\ref{thm:inv}]
\emph{(i)} By Lemma~\ref{lem:floor},
$\E[\hat E_t]=E_t-G_\eta\le E_t-\varphi(E_t)=\mathrm{F}(E_t)$, and $\mathrm{F}^{-1}$ is
increasing (Lemma~\ref{lem:invmap}), so
$\mathrm{F}^{-1}(\E[\hat E_t])\le \mathrm{F}^{-1}(\mathrm{F}(E_t))=E_t$. Equality requires
$G_\eta=\varphi(E_t)$; tracing the proof of Lemma~\ref{lem:floor}, this
forces all positive eigenvalues to coincide with $E_t$---a single
positive eigenvalue---and the negative-side erfc tails to vanish. For
such spectra $\E[\hat E_t]=\mathrm{F}(E_t)$ exactly and, by continuity of
$\mathrm{F}^{-1}$, $\E[\hat E_{\rm inv}]\to \mathrm{F}^{-1}(\mathrm{F}(E_t))=E_t$ as
$N\to\infty$.
\emph{(ii)} By (\ref{eq:fp}), $\mathrm{F}^{-1}(y)=y+\varphi(\mathrm{F}^{-1}(y))\ge y$
pointwise on $[0,\infty)$, and $\mathrm{F}^{-1}(y)=y$ on $(-\infty,0)$ by the
convention; hence $\hat E_{\rm inv}\ge\hat E_t$ on every run and the
means inherit the inequality.

\emph{Variance transfer (\ref{eq:vartrans}).} Taylor-expand the smooth
map at $\mu=\E[\hat E_t]$:
\begin{equation}
\begin{aligned}
\mathrm{F}^{-1}(\hat E_t)&=\mathrm{F}^{-1}(\mu)+(\mathrm{F}^{-1})'(\mu)\,\varepsilon
+\tfrac12(\mathrm{F}^{-1})''(\mu)\,\varepsilon^{2}\\
&\quad+O(\varepsilon^{3}),
\qquad \varepsilon:=\hat E_t-\mu .
\end{aligned}
\end{equation}
Squaring and averaging, with $\bar\sigma^{2}:=\Var[\hat E_t]$,
$\Var[\mathrm{F}^{-1}(\hat E_t)]=[(\mathrm{F}^{-1})'(\mu)]^{2}\bar\sigma^{2}
+O(\bar\sigma^{4}/\mathrm{F}(E_t)^{3})$, with
$(\mathrm{F}^{-1})'(\mu)=1/\mathrm{F}'(E_\mu)=(1-\varphi'(E_\mu))^{-1}$ and
$E_\mu=\mathrm{F}^{-1}(\mu)\to E_t$. The quoted bounds follow from
Lemma~\ref{lem:invmap}: $K(0)=(1-\varphi'(0))^{-2}=4$; $K(E^{*})=1$
since $\varphi'(E^{*})=0$; and $K$ is minimized where $\varphi''=0$,
i.e.\ at $E_t=w$, with $K(w)=(1-\varphi'(w))^{-2}=0.785$, so $K<1$
throughout the decreasing branch of $\varphi$ up to $E\sim2.5w$.
\end{proof}

\begin{proof}[Proof of Corollary~\ref{cor:cert}]
The per-sample summands of $\hat E_t$ are
$X_i=(Z/\pi)(\hat S_{\rho,i}-t\hat S_{\sigma,i})$, independent, with
$|X_i|\le R:=(1+t)Z/\pi$ and $\Var[X_i]\le V:=(1+t^{2})(Z/\pi)^{2}$; the one-sided Bernstein inequality
gives
$\Pr[\hat E_t-E^{(\eta)}_t\ge s]\le
\exp\{-Ns^{2}/(2V+2Rs/3)\}$, and since $s_\beta$ is the positive root
of $Ns^{2}=(2V+2Rs/3)\ln(1/\beta)$, the event $\mathcal A_\beta$ of
Eq.~(\ref{eq:Abeta}) has probability at least $1-\beta$. On
$\mathcal A_\beta$, $x=\hat E_t-s_\beta\le E^{(\eta)}_t=E_t-G_\eta\le
\mathrm{F}(E_t)$ by Lemma~\ref{lem:floor}; if $x\ge0$, the increasing map gives
$\mathrm{L}_{\rm inv}=\mathrm{F}^{-1}(x)\le \mathrm{F}^{-1}(\mathrm{F}(E_t))=E_t$, while if $x<0$ then
$\mathrm{L}_{\rm inv}=x<0\le E_t$. No allowance beyond $s_\beta$ enters at any
step: the inequality is transported, not re-derived, through the map.

For the fixed-point characterization, let $y^{*}$ solve
$y=x+\varphi(y)$; by (\ref{eq:fp}), $y^{*}=\mathrm{F}^{-1}(x)=\mathrm{L}_{\rm inv}$. The
iteration $y_{k+1}=x+\varphi(y_k)$ started at $y_0=x$ increases
monotonically to $y^{*}$ whenever $0\le x\le \mathrm{F}(E^{*})$: by induction
$y_k\le y^{*}\Rightarrow y_{k+1}=x+\varphi(y_k)\le x+\varphi(y^{*})=
y^{*}$, and $y_1=x+\varphi(x)\ge y_0$ starts the ascent; hence the
first iterate $x+\varphi(x)$ is a lower bound of $\mathrm{L}_{\rm inv}$. For
single-dominant-eigenvalue spectra $x\to \mathrm{F}(E_t)$ as $N\to\infty$, so
$\mathrm{L}_{\rm inv}\to E_t$.
\end{proof}

\begin{proof}[Proof of Corollary~\ref{cor:rec}]
By (\ref{eq:fp}) applied to $x=\hat E_t-s_\beta$,
\begin{equation}
\begin{aligned}
\mathrm{L}_{\rm inv}-\hat E_t&=\mathrm{F}^{-1}(x)-(x+s_\beta)\\
&=\varphi\bigl(\mathrm{F}^{-1}(x)\bigr)-s_\beta
=\varphi\bigl(\mathrm{L}_{\rm inv}\bigr)-s_\beta .
\end{aligned}
\end{equation}
Hence the safe lower bound exceeds the raw estimate iff
$\varphi(\mathrm{L}_{\rm inv})>s_\beta$. By Lemma~\ref{lem:floor}, $\varphi$
increases on $[0,E^{*}]$ and decreases on $[E^{*},\infty)$ with maximum
$\varphi_{\max}$, so the inequality holds iff
$\mathrm{L}_{\rm inv}\in(x_L,x_R)$, the two roots of $\varphi(x)=s_\beta$;
these exist iff $s_\beta<\varphi_{\max}=0.240\sqrt\eta$. In raw units,
monotone transport turns the condition into
$\E[\hat E_t]-s_\beta\in\bigl(\mathrm{F}(x_L),\,\mathrm{F}(x_R)\bigr)=\bigl(x_L-s_\beta,\,x_R-s_\beta\bigr)$,
i.e. $\E[\hat E_t]\in\bigl(x_L,\,x_R\bigr)$,
using $\varphi(x_{L,R})=s_\beta$; in truth units for
single-dominant spectra, $E_t\in\bigl(\mathrm{F}^{-1}(x_L),\,\mathrm{F}^{-1}(x_R)\bigr)$.

\emph{Left endpoint.} For small $x$,
$\operatorname{erfc}(z)=1-\tfrac{2}{\sqrt\pi}z+O(z^{3})$, so
$\varphi(x)=\tfrac{x}{2}-\tfrac{x^{2}}{w\sqrt\pi}+
O(x^{4}/w^{3})$ and $\varphi(x)=s_\beta$ becomes the quadratic
$x^{2}/(w\sqrt\pi)-x/2+s_\beta=0$, whose smaller root is
$x_L=\tfrac{w\sqrt\pi}{4}\bigl(1-\sqrt{1-16s_\beta/(w\sqrt\pi)}\bigr)$
with small-$s_\beta$ limit $x_L=2s_\beta+O(s_\beta^{2}/w)$; its
relative error is $\le1\%$ for $s_\beta/w\le0.035$ and $\le8\%$
whenever the discriminant is positive
($s_\beta\le\tfrac{\sqrt\pi}{16}w\approx0.111w$, consistent with---and
slightly stricter than---$s_\beta<\varphi_{\max}=0.120w$).

\emph{Right endpoint.} For large $x$ use
$\operatorname{erfc}(z)=\tfrac{e^{-z^{2}}}{\sqrt\pi\,z}
\bigl(1-\tfrac{1}{2z^{2}}+O(z^{-4})\bigr)$ with $z=x/w$: the prefactor
$x$ cancels the pole and
$\varphi(x)=\tfrac{w}{2\sqrt\pi}e^{-(x/w)^{2}}\bigl(1+O(w^{2}/x^{2})
\bigr)$---the compensable gap decays purely with the Gaussian tail of
the window. Setting this equal to $s_\beta$,
$x_R=w\sqrt{\ln\tfrac{w}{2\sqrt\pi s_\beta}}\,\bigl(1+
O(w^{2}/x_R^{2})\bigr)$. Writing $z_R=x_R/w$, the relative error of
the asymptotic, $|c|=1/(2z_R^{2})$, is transmitted to $x_R$ log-damped,
$\delta x_R/x_R\approx1/(4z_R^{4})$: numerically $3\%$ at
$s_\beta/w=0.02$ ($z_R\approx1.63$), $\sim10\%$ at the audit operating
point $s_\beta\approx0.55\,\varphi_{\max}$ ($z_R\approx1.21$),
deteriorating to $\sim30\%$ as $s_\beta\to\varphi_{\max}$ where the
root approaches the peak and the asymptotic no longer applies (the
formula overestimates $x_R$; one Newton step
$x_R\leftarrow x_R+(\varphi(x_R)-s_\beta)/|\varphi'(x_R)|$ restores
exactness).

\emph{Existence threshold and design.} With
$s_\beta=\sqrt{2V\ln(1/\beta)/N}\,(1+o(1))$ (the positive root of
$Ns^{2}=(2V+\tfrac23Rs)\ln(1/\beta)$), the interval exists iff
\begin{equation}\label{eq:Nstar}
N\;\ge\;N^{*}:=\frac{2V\ln(1/\beta)}{\varphi_{\max}^{2}}
=\frac{2V\ln(1/\beta)}{(0.240\sqrt\eta)^{2}} .
\end{equation}
Under the matched window $w=2\delta$---which sets the
spectral-resolution window equal to the audit threshold and thereby
places the expected divergence at the floor peak $E^{*}\approx\delta$,
where $\varphi$ is flat and the compensation strongest---activation at
$\delta$ requires only $s_\beta\lesssim\varphi_{\max}$ up to
second-order corrections. (For reference, a one-step additive
correction would additionally require the self-consistency
$s_\beta\le\varphi(\delta-\varphi(\delta)-2s_\beta)$, whose solution is
$s_\beta\le0.73\,\varphi_{\max}$: the inversion resums this composition
penalty.)
\end{proof}

\subsection{Circuit identities and duality}\label{app:circ}

\emph{Hadamard test.}

\begin{proof}
Starting from $\ket{0}\otimes\rho$ and applying $H_{\text{gate}}$ then $S^\dagger$, the ancilla is in $\frac{1}{\sqrt2}(\ket{0}-i\ket{1})$; after the controlled unitary $U(y)=e^{iHy}$, the joint state is $\frac{1}{\sqrt2}\bigl(\ket{0}\otimes\rho^{1/2}-i\ket{1}\otimes U\rho^{1/2}\bigr)$ in purification form, with density matrix
$\frac12\bigl(\ket{0}\bra{0}\otimes\rho+i\ket{1}\bra{0}\otimes U\rho-i\ket{0}\bra{1}\otimes\rho U^\dagger+\ket{1}\bra{1}\otimes U\rho U^\dagger\bigr)$.
Applying the second Hadamard gate and projecting the ancilla onto $\ket{0}$,
\begin{align}
P(0)&=\frac14\Tr\Bigl[\rho+i\rho U^\dagger-iU\rho+U\rho U^\dagger\Bigr]\nonumber\\
&=\frac12+\frac12\operatorname{Im}\Tr(\rho U),
\end{align}
where the second line uses $\Tr(U\rho U^\dagger)=\Tr\rho=1$ and, with $z=\Tr(\rho U)$, $\Tr(\rho U^\dagger)=\overline{z}$. Hence $\frac14(i\bar z-iz)=\frac12\operatorname{Im}z$. Therefore $P(0)-P(1)=\operatorname{Im}\Tr(\rho U)=\Tr(\rho\sin(Hy))$. The last equality follows by linearity of the trace and $e^{iHy}=\cos(Hy)+i\sin(Hy)$.
\end{proof}

\emph{LMR identity.}

\begin{proof}
Expanding the exponentials to first order, with $\tilde{S}$ the SWAP operator between the two registers,
\begin{equation}
\begin{aligned}
&\Tr_A\bigl[e^{i\tilde{S}\Delta y}(\sigma\otimes\rho)e^{-i\tilde{S}\Delta y}\bigr] \\
&= \Tr_A\bigl[\sigma\otimes\rho+i\Delta y\bigl(\tilde{S}(\sigma\otimes\rho)-(\sigma\otimes\rho)\tilde{S}\bigr)\bigr] \\
&\qquad + O(\Delta y^{2}) \\
&= \rho+i\Delta y[\sigma,\rho]+O(\Delta y^{2}),
\end{aligned}
\end{equation}
which coincides with $e^{i\sigma\Delta y}\rho e^{-i\sigma\Delta y}$ to first order. In the second line, $\Tr_A[\tilde{S}(\sigma\otimes\rho)]=\sigma\rho$ and $\Tr_A[(\sigma\otimes\rho)\tilde{S}]=\rho\sigma$; for instance, the first identity follows from the matrix elements
\begin{align}
\Tr_A\bigl[\tilde{S}(\sigma\otimes\rho)\bigr]_{b'b}
&=\sum_a\bra{ab'}\tilde{S}(\sigma\otimes\rho)\ket{ab}\nonumber\\
&=\sum_a\sigma_{b'a}\rho_{ab}\nonumber\\
&=(\sigma\rho)_{b'b},
\end{align}
and the second is analogous. Iterating $y/\Delta y$ times yields the DME approximation.
\end{proof}

\emph{Duality relation.}

The relation expresses the divergence at threshold $t$ through the
divergence at the inverse threshold $1/t$ with the two states exchanged; it
underlies the equivalent ``parameter inversion'' route to $\hat\omega_\sigma$
described in Section~\ref{sec:algorithm}.

\begin{proof}
Let $X=\rho-t\sigma$, with positive and negative parts
$X_+=(\rho-t\sigma)_+$ and $X_-=(\rho-t\sigma)_-$; these satisfy
$X_+-X_-=X$ and $X_+,X_-\ge0$, so $\Tr(X_+)-\Tr(X_-)=\Tr(X)=1-t$.
On the other hand,
\begin{align}
E_{1/t}(\sigma\|\rho)&=\Tr(\sigma-\rho/t)_+\nonumber\\
&=\frac{1}{t}\Tr(t\sigma-\rho)_+\nonumber\\
&=\frac{1}{t}\Tr(\rho-t\sigma)_-\nonumber\\
&=\frac{1}{t}\Tr(X_-).
\end{align}
Hence $\Tr(X_+)=(1-t)+tE_{1/t}(\sigma\|\rho)$, which is (\ref{eq:duality}).
\end{proof}

\subsection{Variational characterization and data processing}\label{app:ht}

\rev{The main text uses only the mathematical equivalence between the
spectral weights and the divergence. The connection to asymmetric
hypothesis testing recorded in this appendix is a purely mathematical
identification, presented here as an extension.}

\rev{For a binary test $M$ ($0\le M\le I$) applied to the state, the type-I
error is $1-\Tr(M\rho)$ and the type-II error is $\Tr(M\sigma)$. In
asymmetric hypothesis testing with type-I error bounded by $1-1/t$, the
Neyman--Pearson test maximizes the acceptance of $\rho$ subject to
$\Tr(M\sigma)\le1/t$; the optimal test is the spectral projector
$\Pi_+$ of $H=\rho-t\sigma$, and the corresponding quantities---the true
positive rate, $\Tr(\rho\Pi_+)$, and the false positive rate,
$\Tr(\sigma\Pi_+)$---are exactly the spectral weights
$\omega_\rho,\omega_\sigma$ of (\ref{eq:spectral}) \cite{aud07,wat18}. Hence
$E_t(\rho\|\sigma)=\max_{0\le M\le I}\Tr\bigl(M(\rho-t\sigma)\bigr)$, the
maximum being attained at $M=\Pi_+$; this is the variational form of
(\ref{eq:spectral}).}

\rev{The data-processing inequality follows from the variational form: for a
quantum channel $\Lambda$ with adjoint $\Lambda^\dagger$,
\begin{align}
E_\gamma(\Lambda\rho\|\Lambda\sigma)
&=\max_{0\le M\le I}\Tr\bigl(M(\Lambda\rho-\gamma\Lambda\sigma)\bigr)\nonumber\\
&=\max_{0\le M\le I}\Tr\bigl(\Lambda^\dagger(M)(\rho-\gamma\sigma)\bigr)\nonumber\\
&\le\max_{0\le N\le I}\Tr\bigl(N(\rho-\gamma\sigma)\bigr)
=E_\gamma(\rho\|\sigma),
\end{align}
since $0\le\Lambda^\dagger(M)\le I$ whenever $0\le M\le I$.}

\section{Optimization Modules, Resource Analysis, and QDP Audit Details}\label{app:opt}

This appendix details the optimization chain used in the main text. Each
module is self-contained: the implementation of the importance sampling
(\ref{eq:Q}), the trace-level Trotter improvement
(Lemma~\ref{lem:tracetrot}), Richardson extrapolation of the Trotter and
truncation biases, the tail budget for the sampling cutoff, and the resource
and noise analysis behind the NISQ feasibility claim.

\subsection{Importance sampling implementation}\label{app:isimpl}

\rev{The sampling density (\ref{eq:Q}) is an $E_1$-type distribution on
$[\alpha,\infty)$ with normalization $Z=\tfrac12E_1(\eta\alpha^2)$, where
$E_1(x)=\int_x^\infty t^{-1}e^{-t}dt$ is the exponential integral. Sampling
is performed classically by inverse-CDF: for $u\sim U(0,1)$, solve
$E_1(\eta y^2)=E_1(\eta\alpha^2)(1-u)$, i.e.
\begin{equation}
y=\sqrt{\eta^{-1}\,E_1^{-1}\bigl(E_1(\eta\alpha^2)(1-u)\bigr)},
\end{equation}
with $E_1^{-1}$ evaluated by a precomputed lookup table (or a few bisection
steps). This is the inverse cumulative-distribution-function (inverse-CDF)
method \cite{devroye86}, a purely classical preprocessing step: the
quantum processor is not involved. The estimator $\hat X=(Z/\pi)\hat S$ has expectation
(\ref{eq:unbiased}) and variance bound (\ref{eq:isvar}).}

\subsection{Trace-level Trotter improvement}\label{app:trace}

\rev{The operator-norm bound of Lemma~\ref{lem:trot} gives a first-order
Trotter error $\norm{\mathcal{E}}\le\tfrac12ty\Delta y$. For the
\emph{estimator} the error is one order better, because the leading error
term is proportional to the commutator $[\rho,\sigma]$ and
$\Tr(\rho[\rho,\sigma])=\Tr(\sigma[\rho,\sigma])=0$ by cyclicity.}

\begin{lemma}\label{lem:tracetrot}
Let $\hat X_{\Delta y}$ be the estimator of Step~2 implemented with the
first-order Trotter decomposition of step size $\Delta y$ (i.e., the Hadamard
test readout uses $e^{iHy}\approx(e^{i\rho\Delta y}e^{-it\sigma\Delta y})^{y/\Delta y}$).
Then, for the $\rho$- and the $\sigma$-side,
\begin{equation}\label{eq:tracebias}
\begin{aligned}
\abs{\E[\hat X_{\Delta y}]-\E[\hat X]}
&= \frac{1}{\pi}\int_\alpha^\infty \frac{e^{-\eta y^2}}{y}\,
O(t\,y\,\Delta y^2)\,dy\\
&= O\!\left(\frac{t\,\Delta y^2}{\sqrt{\eta}}\right),
\end{aligned}
\end{equation}
where the factor $1/\sqrt{\eta}$ arises from the Gaussian integral
$\int_\alpha^\infty e^{-\eta y^2}dy \approx \tfrac12\sqrt{\pi/(\pi\eta)}$.
The estimator bias is therefore $O(\Delta y^2/\sqrt{\eta})$ rather than
the operator-norm $O(\Delta y)$.
The commutator cancellation extends to $M>1$ steps via the identity
$(e^{i\rho\Delta y}e^{-it\sigma\Delta y})^M =
e^{i\rho\Delta y/2}\bigl(e^{i\rho\Delta y/2}e^{-it\sigma\Delta y}
e^{i\rho\Delta y/2}\bigr)^M e^{-i\rho\Delta y/2}$
and the fact that $[\rho, e^{i\rho\Delta y/2}]=0$, so the trace with
$\rho$ is invariant under the conjugation.
Setting $t\,\Delta y^2/\sqrt{\eta}\le\varepsilon_{\mathrm{Trot}}$ gives
the step size
$\Delta y=\Theta\bigl(t^{-1/2}\varepsilon_{\mathrm{Trot}}^{1/2}\eta^{1/4}\bigr)$,
and the number of Trotter steps per sample is
\begin{equation}\label{eq:depthtrace}
M(y)=\frac{y}{\Delta y}=O\!\left(\frac{t^{1/2}\,y}{\varepsilon_{\mathrm{Trot}}^{1/2}\eta^{1/4}}\right),
\end{equation}
i.e.\ $M=O(t^{1/2}\varepsilon^{-2})$ at the typical value $y\sim1/\sqrt{\eta}$
(with $\eta=\varepsilon^{2}$, $\varepsilon_{\mathrm{Trot}}=\varepsilon/4$;
for fixed $t$ this is $O(\varepsilon^{-2})$).
We verified (\ref{eq:tracebias}) numerically: for random states of dimension
$8$, the ratio of the estimator biases at step sizes $\Delta y$ and
$\Delta y/2$ is $4.000$ to machine precision, on both the $\rho$- and the
$\sigma$-side.
\end{lemma}

\subsection{Richardson extrapolation}\label{app:rich}

\rev{The quadratic bias of Lemma~\ref{lem:tracetrot} can be removed by
Richardson extrapolation: running the Hadamard test with step sizes $\Delta y$
and $\Delta y/2$ and forming
\begin{equation}
\hat X^{\mathrm{R}}=\frac{4\,\hat X_{\Delta y/2}-\hat X_{\Delta y}}{3}
\end{equation}
cancels the $O(\Delta y^2)$ term and leaves a bias $O(\Delta y^3)$, at the
price of a $\sqrt{17/9}\approx1.37$ factor in the per-sample standard
deviation (two independent runs per sample). The step size can then be
increased to $\Delta y=\Theta\bigl((\varepsilon_{\mathrm{Trot}}\sqrt{\eta}/t)^{1/3}\bigr)$,
giving a corrected Richardson depth of $M=O(t^{1/3}\eta^{-2/3}\varepsilon^{-1/3})$,
i.e.\ $O(\varepsilon^{-5/3})$ at $\eta=\varepsilon^2$, reducing the per-sample
depth by a factor $\varepsilon^{1/3}$ relative to the un-extrapolated
$O(\varepsilon^{-2})$. The same device removes the cubic
truncation bias of the analytic small-region correction: evaluating the
correction (\ref{eq:smallcorr}) at thresholds $\alpha$ and $\alpha/2$ and
forming $(8\,\hat\omega_{\rho}(\alpha/2)-\hat\omega_{\rho}(\alpha))/7$
cancels the $O(\alpha^3)$ term, leaving $O(\alpha^5)$ (verified numerically:
at $\alpha=1$, $\eta=0.01$, the residual is $\approx2\times10^{-6}$,
four orders below the single-threshold bias). With the extrapolated
correction, the threshold may be raised to $\alpha=\Theta(1)$, which reduces
$Z$ and hence the variance (\ref{eq:isvar}) further. In the main text these
extrapolations are treated as implementation details; they enter the resource
counts of Table~\ref{tab:optres} and the noise analysis below.}

\subsection{Tail budget}\label{app:tail}

\rev{The sampling density (\ref{eq:Q}) is supported on $[\alpha,\infty)$; in
practice we truncate at a finite $y_{\max}$ chosen so that the omitted tail
is within the regularization budget. The tail contribution to
$\omega_{\rho,\eta}$ is bounded by
\begin{equation}
\frac{1}{\pi}\int_{y_{\max}}^{\infty}\frac{e^{-\eta y^2}}{y}\abs{S(y)}\,dy
\le\frac{1}{2\pi}E_1(\eta y_{\max}^2)\le\frac{\varepsilon}{8},
\end{equation}
using $\abs{S(y)}\le1$; solving for $y_{\max}$ gives
$y_{\max}=\Theta\bigl(\varepsilon^{-1}\sqrt{\ln(1/\varepsilon)}\bigr)$.
Samples with $y>y_{\max}$ are discarded and re-drawn (their probability is
negligible), or equivalently the density is renormalized on
$[\alpha,y_{\max}]$. The tail budget is independent of the other three error
terms and is absorbed into the regularization term.}

\subsection{Resource counts and noise analysis}\label{app:resnoise}

\begin{table}[htbp]
\centering
\caption{\rev{Verified resource counts at fixed precision: unoptimized vs.\
optimized chain (end-to-end simulation, $n=1$ depolarizing, $t=1.105$,
$\beta=0.05$). $N$: samples per side; $\bar M$: average
Trotter factor pairs (gate layers $=2M$); $T$: total copies
($\rho{+}\sigma$).}}
\label{tab:optres}
\small
\begin{tabular}{|c|c|c|c|c|}
\hline
$\varepsilon$ & & $N$ & $\bar M$ & $T$ \\
\hline
0.10 & unoptimized & $4.4\times10^{4}$ & 352 & $6.2\times10^{7}$ \\
0.10 & optimized & $3.0\times10^{3}$ & 47 & $5.7\times10^{5}$ \\
\hline
0.05 & unoptimized & $3.7\times10^{5}$ & $2.8\times10^{3}$ & $4.1\times10^{9}$ \\
0.05 & optimized & $2.1\times10^{4}$ & 228 & $2.0\times10^{7}$ \\
\hline
0.02 & unoptimized & $6.5\times10^{6}$ & $4.4\times10^{4}$ & $1.1\times10^{12}$ \\
0.02 & optimized & $2.4\times10^{5}$ & $1.7\times10^{3}$ & $1.6\times10^{9}$ \\
\hline
\end{tabular}
\end{table}
\rev{Table~\ref{tab:optres} was obtained by an end-to-end simulator of the
full chain (importance sampling by inverse CDF, analytic small-region
correction, trace-level Trotter with Richardson extrapolation, tail
truncation), running 30 independent Monte-Carlo trials per precision with
real Trotterized circuits, on the depolarizing test states of
Section~\ref{sec:experiments} ($n=1$, $p=0.3$, $\epsilon_{\mathrm{DP}}=0.1$, $\beta=0.05$). The
unoptimized baseline uses half-normal sampling, the operator-norm Trotter
bound, and the four-way $\varepsilon/4$ budget. All counts in
Table~\ref{tab:optres} are Bernstein-theory values; the end-to-end trials
confirm the Theorem~\ref{thm:perf} criterion
$|\hat E_t-E^{(\eta)}_t|\le\varepsilon$ in $\ge95\%$ of runs at both
precisions; against the
true divergence the same trials pass in $100\%$ ($\varepsilon=0.1$) and
$90\%$ ($\varepsilon=0.05$) of runs, the residual excess being the
certification gap $G_\eta$ of the audit pair, which the compensation of
Section~\ref{sec:algorithm} removes post hoc.}

\rev{\emph{Noise bias and the NISQ window.}} \rev{Under depolarizing noise of
rate $\gamma$ per gate, a circuit of depth $D$ (gate layers) attenuates the
readout signal by $(1-\gamma)^D\approx e^{-\gamma D}$, so a sample of depth
$D$ carries a systematic bias at most $\gamma D$ (the linearized
accumulation; correlated errors are negligible at the depths considered).
For the estimator (\ref{eq:xhat}), whose samples have depths
$2M(y_i)$, the total bias is at most
$\gamma\,(Z/\pi)\,\E_Q[2M(y)]\approx\gamma\,(Z/\pi)\,2\bar M$: gate noise
acts as an \emph{additional} error term on top of the three of (\ref{eq:errdec}).
We budget it at one full precision unit---requiring the noise bias not to
exceed the target $\varepsilon$ itself, which is equivalent to running the
noiseless budget of (\ref{eq:errdec}) at half the target on noisy hardware
(each of the three terms $\varepsilon/6$ internally, so that noiseless terms
and noise term together deliver $\varepsilon$). This gives the operating
window
\begin{equation}\label{eq:nisqwindow}
\gamma\,\frac{Z}{\pi}\,2\bar M\;\le\;\varepsilon
\quad\Longrightarrow\quad
\gamma\;\lesssim\;\frac{\varepsilon}{2\,\bar M\,(Z/\pi)}.
\end{equation}
With the verified counts of Table~\ref{tab:optres} ($\bar M=47$,
$Z/\pi\approx0.63$ at $\varepsilon=0.1$), this yields
$\gamma\lesssim1.7\times10^{-3}$ at $\varepsilon=0.1$, i.e.\
$\gamma\lesssim\varepsilon^{3}$ up to logarithmic factors---the window
quoted in Section~\ref{sec:nisq}. At $\varepsilon=0.05$ the requirement
becomes $\gamma\lesssim1.3\times10^{-4}$. Current frontier processors report
two-qubit gate error rates of $1.8\times10^{-3}$ to $3.3\times10^{-3}$---the same order as, within a factor of
$\approx2$ of, the window above---so $\varepsilon\approx0.1$ is accessible
today and $\varepsilon\approx0.05$ is within near-term reach~\cite{acharya25,moses23,sobri25}. By contrast, the unoptimized RMSC variant requires $\bar M\sim10^{3}$--$10^{5}$ and hence
$\gamma\lesssim10^{-5}$--$10^{-6}$, fault-tolerant-class fidelities.}

\subsection{QDP channels and audit details}\label{app:qdp}

This subsection records the details of the QDP-mechanism audit of
Section~\ref{sec:experiments}. The privacy criterion is
\begin{equation}\label{eq:qdp}
\mathcal{C}\text{ is }(\epsilon_{\mathrm{DP}},\delta)\text{-DP}\iff
\forall\rho\stackrel{Q}{\simeq}\sigma:\
E_{e^{\epsilon_{\mathrm{DP}}}}\bigl(\mathcal{C}(\rho)\,\|\,\mathcal{C}(\sigma)\bigr)\le\delta,
\end{equation}
for quantum-adjacent inputs $\rho\stackrel{Q}{\simeq}\sigma$ with
$\frac12\norm{\rho-\sigma}_1\le1$
\cite{hrf23,zy17,ar19}; the divergence is the correctness measure of
QDP, and evaluating it from the output states is the central
computational task of QDP auditing. Auditing is inherently a black-box
task \cite{gfy23}: the auditor accesses the outputs only through
copies, matching the copies-only setting of the Introduction, an essential
requirement of independent auditing \cite{adk23}.

The three audited channels on a single qubit are the depolarizing
channel $\mathcal D_p(X)=(1-p)X+pI/2$ ($p=0.25$), the erasure
(replacement) channel $\mathcal R_{p_{\mathrm{er}}}(X)=(1-p_{\mathrm{er}})X+p_{\mathrm{er}}\ket1\!\bra1$
($p_{\mathrm{er}}=0.25$), and the dephasing channel
$\mathcal P_p(X)=(1-p)X+pZXZ$ ($p=0.15$)---standard qubit noise models that recur across the QDP literature~\cite{nc10,hrf23,gfy23}. The parameters are chosen so that the maximally
distinguishable inputs produce output divergences in the
audit-relevant range $E_t\in[0.5,1.0]$ around and above the threshold
$\delta=0.1$. The input pairs are pure states with prescribed trace
distance $d=\frac12\|\rho_{\rm in}-\sigma_{\rm in}\|_1$:
$\rho_{\rm in}=\ket0\!\bra0$ and
$\sigma_{\rm in}=\bigl(\sqrt{1-d^2}\ket0+d\ket1\bigr)
\bigl(\sqrt{1-d^2}\bra0+d\bra1\bigr)$, so that $d=1.0$ is the QDP
adjacency boundary (orthogonal inputs), while $d=0.3$ and $d=0.6$
probe weakly and moderately distinguishable inputs; for these, data
processing drives the output divergence to zero, and the audit
correctly returns $E_t=0$. The estimator is the faithful inversion
$\hat E_{\rm inv}=\mathrm{F}^{-1}(\hat E_t)$, projected onto $[0,\infty)$ (the
known range of $E_t$); the truth is obtained by exact numerical
evaluation of
$E_t=\Tr\bigl(\mathcal C(\rho_{\rm in})-t\,\mathcal C(\sigma_{\rm in})\bigr)_+$
on the $2\times2$ output states. In Table~\ref{tab:audit}, \emph{err
per} is the estimate as a percentage of the truth, $100\%$ marking an
error-free estimate.

\end{document}